\documentclass{pasj01}
\Received{2019 January 8}
\Accepted{2019 August 14}
\Published{$\langle$publication date$\rangle$}

\begin{document} 

\title{ 
Outer rotation curve of the Galaxy with VERA IV: Astrometry of IRAS$\,$01123+6430 and the possibility of cloud-cloud collision}

\author{
Nagito \textsc{Koide}\altaffilmark{1},%
Hiroyuki \textsc{Nakanishi}\altaffilmark{1},
Nobuyuki \textsc{Sakai}\altaffilmark{2,3},
Asao \textsc{Habe}\altaffilmark{4},
Kazuhiro \textsc{Shima}\altaffilmark{5},
Tomoharu \textsc{Kurayama}\altaffilmark{6},
Mitsuhiro \textsc{Matsuo}\altaffilmark{1,7},
Daisuke \textsc{Tezuka}\altaffilmark{1},
Kohei \textsc{Kurahara}\altaffilmark{1},
Saeko \textsc{Ueno}\altaffilmark{1},
Ross A. \textsc{Burns}\altaffilmark{1,2,3,8},
Akiharu \textsc{Nakagawa}\altaffilmark{1},
Mareki \textsc{Honma}\altaffilmark{2,9},
Katsunori \textsc{M. Shibata}\altaffilmark{2,9},
Takumi \textsc{Nagayama}\altaffilmark{2},
and
Noriyuki \textsc{Kawaguchi}\altaffilmark{2,9}
}%
 \altaffiltext{1}{Graduate School of Science and Engineering, Kagoshima
   University, 1-21-35 Korimoto Kagoshima}
 \email{k1847481@kadai.jp}
 \altaffiltext{2}{Mizusawa VLBI Observatory, National Astronomical
   Observatory of Japan, 2-21-1 Osawa, Mitaka, Tokyo 181-8588, Japan}
 \altaffiltext{3}{Korea Astronomy and Space Science Institute, 776 Daedeokdae-ro, Yuseong-gu, Daejeon 34055, Republic of Korea}
   \altaffiltext{4}{Astrophysics Group, Division of Physics, Graduate School of Science, Hokkaido University Sapporo, Japan} 
 \altaffiltext{5}{Theoretical Astrophysics Group, Department of Physics, Kyoto University, Sakyo-ku, Kyoto 606-8502, Japan}
 \altaffiltext{6}{Center for Fundamental Education, Teikyo University of Science, 2525 Yatsusawa, Uenohara, Yamanashi 409-0193}
 \altaffiltext{7}{Nobeyama Radio Observatory, National Astronomical Observatory of Japan, National Institutes of Natural Sciences, 462-2 Nobeyama Minamimaki, Minamisaku, Nagano 384-1305, Japan}
 \altaffiltext{8}{Joint Institute for VLBI ERIC, Oude Hoogeveensedijk 4, 7991 PD Dwingeloo, The Netherlands}
 \altaffiltext{9}{Department of Astronomical Science, The Graduate University for Advanced Studies, Mitaka, 181-8588}


\maketitle

\begin{abstract}
\ As part our investigation into the Galactic rotation curve, we carried out Very Long Baseline Interferometry (VLBI) observations towards the star-forming region IRAS$\,$01123+6430 using VLBI Exploration of Radio Astrometry (VERA) to measure its annual parallax and proper motion. The annual parallax was measured to be $0.151\,\pm\,0.042\>{\rm mas}$, which corresponds to a distance of $D\ =\ 6.61^{+2.55}_{-1.44}\>{\rm kpc}$, and the obtained proper motion components were $(\mu_\alpha{\rm cos}\delta,\ \mu_\delta)\ =\ (-1.44\,\pm\,0.15,\ -0.27\,\pm\,0.16)\>{\rm mas\>yr^{-1}}$ in equatorial coordinates. Assuming Galactic constants of $(R_0,\ \Theta_0)\,=\,(8.05\,\pm\,0.45\>{\rm kpc},\ 238\,\pm\,14\>{\rm km\>s^{-1}})$, the Galactocentric distance and rotation velocity were measured to be $(R,\ \Theta)\ =\ (13.04\,\pm\,2.24\>{\rm kpc},\ 239\,\pm\,22\>{\rm km\>s^{-1}})$, which are consistent with a flat Galactic rotation curve. The newly estimated distance provides a more accurate bolometric luminosity of the central young stellar object, $L_{\rm Bol}\ =\ (3.11\,\pm\,2.86)\ \times\ 10^3\>L_{\odot}$, which corresponds to a spectral type of B1-B2. The analysis of $\atom{CO}{}{12}$ $(J\,=\,1$--$0)$ survey data obtained with the Five College Radio Astronomical Observatory (FCRAO) 14$\>$m telescope shows that the molecular cloud associated with IRAS$\,$01123+6430 consists of arc-like and linear components, which well matches a structure predicted by numerical simulation of the cloud-cloud collision (CCC) phenomenon. The coexistence of arc-like and linear components implies that the relative velocity of initial two clouds was as slow as $3$--$5\>{\rm km\ s^{-1}}$, which meets the expected criteria of massive star formation where the core mass is effectively increased in the presence of low relative velocity ($\sim 3$--$5\>{\rm km\>s^{-1}}$), as suggested by \citet{2014ApJ...792...63T}.\end{abstract}

\section{Introduction}

The rotation velocity is one of the fundamental parameters concerning the study of the dynamics of the Galaxy (e.g. \cite{2017PASJ...69R...1S}). While it can be relatively easily derived within the solar circle $\timeform{270D}\ <\ l\ <\ \timeform{90D}$, using the terminal velocity (\cite{2004ApJ...607L.127M}; \cite{2006ApJ...638..196M}), beyond the solar circle $\timeform{90D}\ \leq\ l\ \leq\ \timeform{270D}$ it needs to be derived by measuring distances and velocities for Galactic objects.

The Outer Rotation Curve (ORC) project is one of the major science projects carried out with VLBI Exploration of Radio Astrometry (VERA), which comprises four 20$\>$m diameter antennas located at Mizusawa, Ogasawara, Iriki, and Ishigaki, being operated by National Astronomical Observatory of Japan (NAOJ) and Kagoshima University. The aim of the ORC project is to measure the Galactic rotational velocities via parallax measurements for individual 22$\>$GHz ${\rm H_2O}$ masers associated with star forming regions in the Outer Galaxy (\cite{2012PASJ...64..108S}; \cite{2015PASJ...67...68N}; \cite{2015PASJ...67...69S}).

IRAS$\,$01123+6430 is an ORC object and is a high mass star forming region, located at $(\alpha,\ \delta)\ =\ (\timeform{1h15m40.8s},\ \timeform{+64D46^{\prime}40.8^{\prime\prime}})$ (J2000) in equatorial coordinates and $(l, b)=(\timeform{125.51D}, \timeform{+2.03D})$ in Galactic coordinates (\cite{1990A&AS...84..179C}; \cite{1993A&AS..101..153P}). $\atom{CO}{}{}$ molecular lines $\atom{CO}{}{12}$ $(J\,=\,2$--$1)$ and $\atom{CO}{}{12}(J=3-2)$ were detected towards this object based on observations conducted with the KOSMA 3$\,$m telescope \citep{1993A&AS...98..589W}. The $\atom{CS}{}{12}$ $(J\,=\,2$--$1)$ line was also detected by observations with the Onsala 20$\,$m radio telescope \citep{1998A&AS..133..337Z}, indicating the presence of dense gas.

As Part IV of the ORC project series, we report on the measurement of the annual parallax and proper motion of IRAS$\,$01123+6430 and discuss its Galactic rotational velocity. Using our distance measurement we place constraints on the spectral type of the central young stellar object, and discuss the physical properties of the associated molecular cloud. We describe the observation of IRAS$\,$01123+6430 with VERA in section 2, data analysis in section 3, and the resultant parallax, distance, kinematics, physical quantities, and Galactic location of IRAS$\,$01123+6430 in section 4. In section 5, we discuss a possible scenario of high-mass star formation around IRAS$\,$01123+6430 using archival data of Five College Radio Astronomical Observatory (FCRAO). Finally, we summarize the paper in section 6.

\begin{table*}
  \caption{Summary of observations$^{\ast}$}\label{sum_obs}
  \begin{center}
    \begin{tabular}{cccrrrrrrr}
      \hline
      \hline
Epoch & Date        & DOY & \multicolumn{4}{c}{$T_{\rm sys}\>{\rm [K]}$} & Synthesized beam  & PA  & $N_{\rm spot}$ \\
\cline{4-7}
      &             &  [days]  & MIZ & IRK & OGA & ISG & (mas $\times$ mas) & ($\deg$) \\
      \hline
r12087a (1)     & 2012 MAR 27 & 87 & 336.7 & 141.7 & 173.1 & 185 & 1.17 $\times$ 0.79 & $-37$        &    25     \\
r12135c (2)     & 2012 MAY 14 & 135 & 954.7 & 811.6 & 325.1 & 511.3 & 1.18 $\times$ 0.77 & $-32$       &    22      \\
r12234b (3)     & 2012 AUG 21 & 234 & 462.9 & 428.1 & 544.9 & 1009.9 & 1.18 $\times$ 0.79 & $-40$      &    22      \\
r13025a (4)     & 2013 JAN 25 & 390 & 190.2 & 134.8 & 303.5 & 286.9 & 1.22 $\times$ 0.79 & $-50$       &     14     \\
r13069a (5)     & 2013 MAR 10 & 434 & 119.1 & 216.6 & 176.5 & 231.6 & 1.19 $\times$ 0.78 & $-55$      &     13       \\
r13227c (6)     & 2013 AUG 15 & 592 & 291.3 & 598.4 & 546.3 & 1851.1 & 1.27 $\times$ 0.75 & $-59$      &     14       \\
r13292b (7)     & 2013 OCT 19 & 657 & 214.1 & 287.5 & 426.1 & 343.3 & 1.15 $\times$ 0.75 & $-47$      &     12     \\
r14082a (8)     & 2014 MAR 23 & 812 & 129.2 & 149.3 & 179.8 & 231.8 & 1.19 $\times$ 0.78 & $-50$     &     13       \\
r14144b (9)     & 2014 MAY 24 & 874 & 224.8 & 215.7 & 195.6 & 433.8 & 1.19 $\times$ 0.79 & $-48$     &    11        \\
r14245b (A)     & 2014 SEP 02 & 975 & 233.6 & 1709.2 & 748.2 & 863.1 & 1.17 $\times$ 0.77 & $-48$     &    5        \\
r14314a (B)     & 2014 NOV 10 & 1044 & 149.9 & 234.5 & 899.6 & 275.7 & 1.26 $\times$ 0.73 & $-54$  &  10             \\
      \hline
     \hline
    \end{tabular}
  \end{center}
{$^\ast$ DOY (Day-of-year) indicates days from 2012 January 1. $T_{\rm sys}$ at four station is following: Mizusawa (MIZ), Iriki (IRK), Ogasawara (OGA), and Ishigaki (ISG). PA shows the position angle east of north for the maser spot. $N_{\rm spot}$ shows the number of detected maser spot.}
\end{table*}

\section{Observation}

IRAS$\,$01123+6430 was observed at the ${\rm H_2O}$ $6_{16}\rightarrow 5_{23}$ transition line (rest frequency of 22.235080$\>$GHz) with VERA from 2012 January 20 to 2014 November 10. Though we obtained data in 13$\>$epochs, data of the first epoch and of DOY (day of year) 355 in 2013 were not used. This is because the tracking centre was modified after the first epoch so that maser emission can be detected near the tracking centre and the latter observation failed due to critical malfunctions at two stations. We observed the target object, IRAS$\,$01123+6430, and the distant quasar J$\,$0128+63, as an astrometric phase reference source simultaneously in dual beam mode to correct for tropospheric phase fluctuation \citep{2000SPIE.4015..544K}. This approach improves the signal-to-noise ratio of the target image with phase referencing and allows us to measure the accurate position of the target, relative to the reference source. The beams pointed to IRAS$\,$01123+6430 and J$\,$0128+63 are referred to as the A- and B-beams, respectively in this paper. The separation between these two objects was $\timeform{2.18D}$. In addition, we observed one of DA$\,$55, 3$\,$C$\,$84, 3$\,$C$\,$454.3, or OJ$\,$287 every 80$\>$minutes as clock drift calibrators and for bandpass and fringe fitting.

The data obtained before 2014 September were recorded to magnetic tape using the DIR$\,$2000 recorder at 1024$\>$Mbps, while the new HDD recording system "OCTADISK" \citep{2012ivs..conf...91O} was used to record data at 1024$\>$Mbps after 2014 September. These data were sent to Mizusawa VLBI observatory for correlation using a software correlator.

The received signal in the observations was separated into 16 intermediate frequency (IF) channels after digital filtering \citep{2005PASJ...57..259I}, each of which had a bandwidth of 16$\>$MHz. One of the IF channels with a frequency matching the water maser transition was assigned to A-beam and the others were assigned to B-beam. The recorded time-domain data were transformed into cross power spectra with 512$\>$channels for the A-beam and with 64 channels for the B-beam, respectively by fast Fourier transformation (FFT). The channel separation of the A-beam was 31.25$\>$kHz, which corresponded to a velocity spacing of $0.42\>{\rm km\ s^{-1}}$. The channel separation of the B-beam was 250$\>$kHz for continuum sources.
 
Parameters such as the observation date, typical system temperature ($T_{\rm sys}$), the size of synthesized beam, its position angle, and the number of detected maser spots are summarized in table \ref{sum_obs}.



\section{Data reduction}

We analyzed the observational data of IRAS$\,$01123+6430 with the Astronomical Image Processing System (AIPS) following the procedure described in \citet{2011PASJ...63..513K}.

\begin{figure*}
\begin{center}
\FigureFile(170mm,170mm){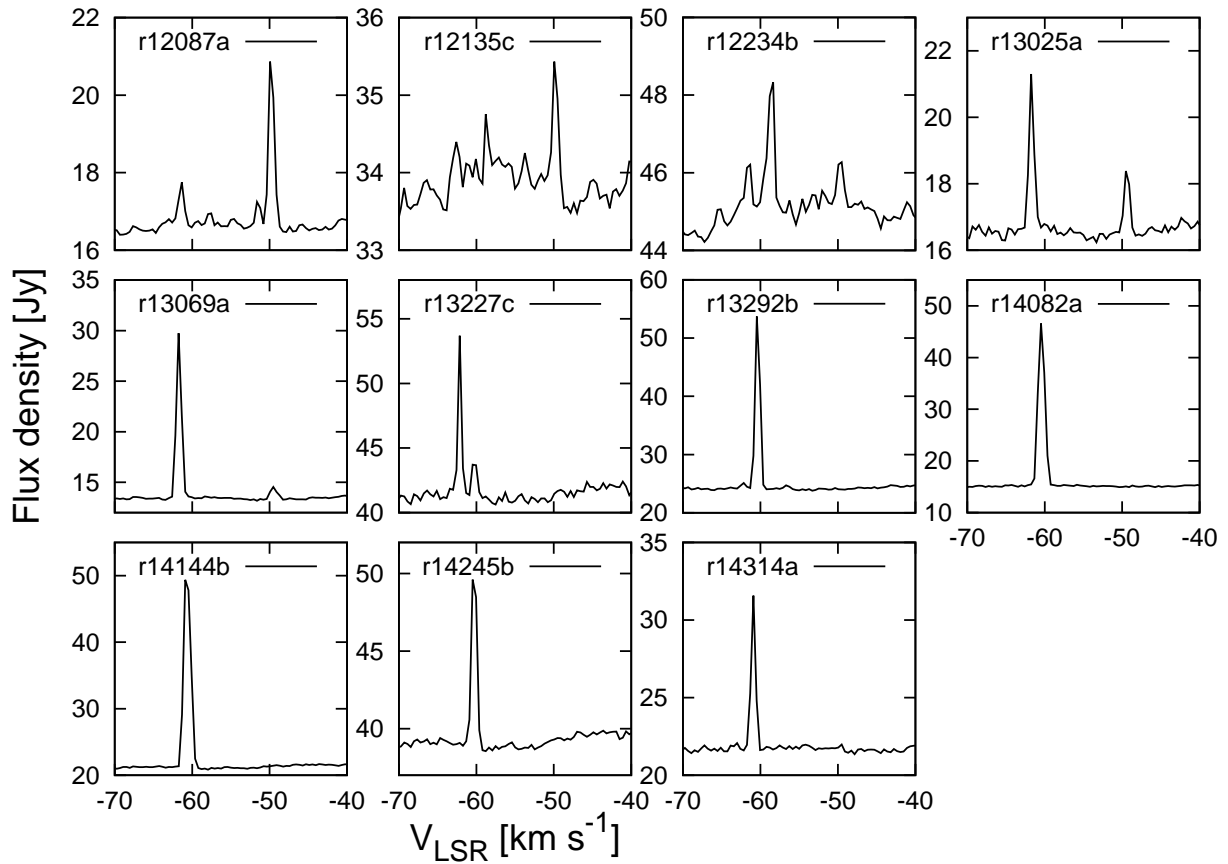}
\end{center}
\caption{Scalar averaged cross power spectra of water maser emission in IRAS$\,$01123+6430 for all epochs. The baselines of these spectra are "MIZ-IRK."}\label{power_spectrum}
\end{figure*}

The data reduction was conducted as follows: (1) loading A-beam fits data and B-beam fits data (task FITLD); (2) dividing cross correlation by auto correlation, to normalize the visibility (ACCOR); (3) converting the units of the amplitude of the visibility data to Jy (APCAL) based on the System Equivalent Flux Density (SEFD) information for VERA stations; (4) loading delay-recalculation data for applying the latest station positions, tropospheric and ionospheric calibrations, and flagging outlier data (TBIN and SNEDT); (5) fringe searching for the B-beam data and applying the result to A-beam data for the phase referencing (FRING, TACOP, and CLCAL); (6) calibrating the reference source by iterative self-calibration (IMAGR, CALIB, CLCAL); (7) applying the result of self-calibration to A-beam data (TACOP, CLCAL); (8) loading dual-beam phase calibration data obtained by the horn-on-dish method, where artificial noise sources mounted on the antenna feed were used to monitor the instrumental delay \citep{2008PASJ...60..935H} and applying the solutions to the A-beam data (TBIN, CLCAL); (9) setting the LSR (local standard of rest) velocity of IRAS$\,$01123+6430 ($V_{\rm{LSR}}\ =\ -55.0\>{\rm km\>s^{-1}}$) to the central channel (256th) of the IF band. And correcting the doppler shift due to the earth's rotation and its motion within the Solar System and towards the LSR (SETJY and CVEL). Throughout this procedure, maser emission peaks were detected in the velocity ranges of $V_{\rm LSR}\ =\ -49.5$ to $-48.7\>{\rm km\>s^{-1}}$ (241--243 channels), $V_{\rm LSR}\ =\ -62.2$ to $-60.9\>{\rm km\ s^{-1}}$ (270--273 channels), and $V_{\rm{LSR}}\ =\ -60.9$ to $-59.6\>{\rm km\>s^{-1}}$ (267--270$\>$channels) as shown in figure \ref{power_spectrum}. 

Next, we made dirty maps of maser spots in the channel range of 230 to 280, and performed CLEANing to improve the S/N ratio of the map (IMAGR). Finally, we derived the positions, the LSR velocities, and the intensities of maser emission by fitting 2-D Gaussian models to the data (JMFIT) and made a list of detected maser spots.

\section{Results}

We detected 5 to 25 maser spots in each epoch as shown in table \ref{sum_obs}. The list of all detected spots and features is shown in table \ref{spotlist}. If a "spot" has a position within a beam size with another spot and velocity channels between them are successive, these spots were identified as the same "feature." We also identified the same maser features between consecutive epochs by checking if the maser feature is detected in the same channel and if the estimated proper motion is within a reasonable range considering the typical Galactic rotational speed of $220\>{\rm km\ s^{-1}}$. Finally, six maser features were identified, and indexed as features 1, 2, 3, 4, 5 and 6. All identified maser features are listed in table \ref{spotlist_1}. It should be noted that features 3 and 4 are spatially overlapped though they have different radial velocities. We found that features 3 and 4 have different proper motions and concluded that these should be identified as two different features.

\begin{table*}
  \caption{The spot list of IRAS$\,$01123+6430 observed in ${\rm H_2O}$ maser line emission$^{\ast}$}\label{spotlist}
  \begin{center}
    \begin{tabular}{cccrrrrrrcr}
    \hline
    DOY & ID & Feature & Intensity & error & \multicolumn{4}{c}{Offset} & Channel & $V_{\rm LSR}$\\
    \cline{6-9}
    & &  & &  & R.A. & error & Decl. & error &  & \\
     \hline
    & &  & $[{\rm Jy\>beam^{-1}}]$ & &  [mas] &  & [mas] &  & & $[{\rm km\>s^{-1}}]$\\
     \hline\hline
87 & 1 & 1 & $1.550$ & $0.056$ & $0.054$ & $1.580\ \times\ 10^{-2}$ & $-0.401$ & $1.699\ \times\ 10^{-2}$ & 241 & $-48.7$\\
87 & 2 & 1 & $4.085$ & $0.112$ & $0.085$ & $1.227\ \times\ 10^{-2}$ & $-0.449$ & $1.307\ \times\ 10^{-2}$ & 242 & $-49.1$\\
87 & 3 & 1 & $8.010$ & $0.205$ & $0.121$ & $1.102\ \times\ 10^{-2}$ & $-0.498$ & $1.164\ \times\ 10^{-2}$ & 243 & $-49.5$\\
87 & 4 & 2 & $9.049$ & $0.232$ & $0.115$ & $1.125\ \times\ 10^{-2}$ & $-0.492$ & $1.166\ \times\ 10^{-2}$ & 244 & $-49.9$\\
87 & 5 & 2 & $3.577$ & $0.104$ & $0.141$ & $1.254\ \times\ 10^{-2}$ & $-0.520$ & $1.325\ \times\ 10^{-2}$ & 245 & $-50.4$ \\
87 & 19 & 3 & $2.376$ & $0.078$ & $3.294$ & $1.443\ \times\ 10^{-2}$ & $-2.657$ & $1.364\ \times\ 10^{-2}$ & 270 & $-60.9$\\
87 & 20 & 3 & $4.414$ & $0.122$ & $3.326$ & $1.151\ \times\ 10^{-2}$ & $-2.656$ & $1.201\ \times\ 10^{-2}$ & 271 & $-61.3$\\
87 & 21 & 4 & $3.455$ & $0.092$ & $3.311$ & $1.171\ \times\ 10^{-2}$ & $-2.672$ & $1.182\ \times\ 10^{-2}$ & 272 & $-61.7$\\
87 & 22 & 4 & $2.141$ & $0.066$ & $3.248$ & $1.331\ \times\ 10^{-2}$ & $-2.693$ & $1.405\ \times\ 10^{-2}$ & 273 & $-62.2$\\
135 & 26 & 1 & $0.836$ & $0.123$ & $-0.093$ & $4.798\ \times\ 10^{-2}$ & $-0.380$ & $7.806\ \times\ 10^{-2}$ & 241 & $-48.7$\\
135 & 27 & 1 & $2.783$ & $0.135$ & $-0.059$ & $1.921\ \times\ 10^{-2}$ & $-0.419$ & $2.817\ \times\ 10^{-2}$ & 242 & $-49.1$\\
135 & 28 & 1 & $5.791$ & $0.211$ & $-0.020$ & $1.330\ \times\ 10^{-2}$ & $-0.443$ & $1.884\ \times\ 10^{-2}$ & 243 & $-49.5$\\
135 & 29 & 2 & $6.929$ & $0.245$ & $-0.031$ & $1.369\ \times\ 10^{-2}$ & $-0.468$ & $2.066\ \times\ 10^{-2}$ & 244 & $-49.9$\\
135 & 30 & 2 & $2.907$ & $0.144$ & $0.001$ & $1.958\ \times\ 10^{-2}$ & $-0.490$ & $3.000\ \times\ 10^{-2}$ & 245 & $-50.4$\\     
135 & 42 & 3 & $2.407$ & $0.144$ & $3.132$ & $2.323\ \times\ 10^{-2}$ & $-2.585$ & $2.775\ \times\ 10^{-2}$ & 270 & $-60.9$\\
135 & 43 & 3 & $3.459$ & $0.155$ & $3.209$ & $1.546\ \times\ 10^{-2}$ & $-2.653$ & $2.021\ \times\ 10^{-2}$ & 271 & $-61.3$\\
135 & 44 & 4 & $2.362$ & $0.133$ & $3.167$ & $1.855\ \times\ 10^{-2}$ & $-2.681$ & $2.616\ \times\ 10^{-2}$ & 272 & $-61.7$\\
135 & 45 & 4 & $2.381$ & $0.147$ & $3.141$ & $2.145\ \times\ 10^{-2}$ & $-2.668$ & $3.198\ \times\ 10^{-2}$ & 273 & $-62.2$\\
     \hline
      \hline
    \end{tabular}
  \end{center}
   {$^{\ast}$The printed version contains only a part of the entire spots list. The complete list is available in the electronic version.}
\end{table*}

\begin{figure*}
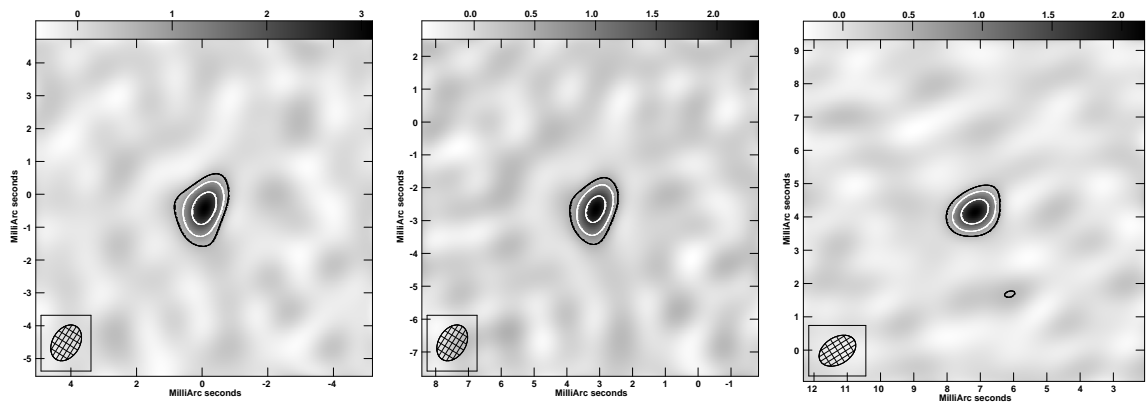

\begin{center}
\FigureFile(50mm,50mm){figure/R12135C_A_CH241_243_AV_CLEAN_NOLAB.PS}
\FigureFile(50mm,50mm){figure/R12135C_A_CH272_273_AV_CLEAN_NOLAB.PS}
\FigureFile(50mm,50mm){figure/R13069A_A_CH269_CLEAN_NOLAB.PS}
\end{center}
\caption{Phase-referenced images for maser features 1, 4 and 6 from left to right (see table \ref{spotlist_1}). $1\sigma$ image noise levels are $0.17\>{\rm Jy\>beam^{-1}}$, $0.15\>{\rm Jy\>beam^{-1}}$, and $0.12\>{\rm Jy\>beam^{-1}}$ from left to right. Contours are drawn by $-3\sigma$, $3\sigma$, $6\sigma$, $12\sigma$, $24\sigma$, and $48\sigma$. (Left) The maser spot of feature 1 (averaged for $V_{\rm{LSR}}\ =\ -49.5$ to $-48.7\>{\rm km\>s^{-1}}$) is shown for r12135c epoch. (Middle) Same as (left), but for feature 4 (averaged for $V_{\rm{LSR}}\ =\ -62.2$ to $-61.7\>{\rm km\>s^{-1}}$) and for r12135c epoch. (Right) Same as (left), but for feature 6 (only for $V_{\rm{LSR}}\ =\ -60.5\>{\rm km\>s^{-1}}$) and for r13069a epoch.}
\label{spots_image_f146}
\end{figure*}

Using multi-epoch astrometric data we measured the annual parallaxes for only features 1, 4 and 6 as shown in table \ref{feat_146}, though features 2, 3 and 5 were not included in the analysis because they were not detected longer than a year. Table \ref{feat_146} shows measured annual parallaxes for the cases of individual maser spots such as spots 1a--1c, 4a--4b, 6a--6d, individual maser features such as features 1, 4, 6, and combined fitting using all spots. We show phase-referenced images for features 1, 4 and 6 in figure \ref{spots_image_f146}.

\begin{table*}
  \caption{The list of identified maser features$^{\ast}$}\label{spotlist_1}
  \begin{center}
    \begin{tabular}{llrrrrl}
      \hline
      \hline
Feature & ID & $\alpha_0$ & $\delta_0$ & Intensity & $V_{\rm LSR}$ & Epochs\\
 &  & $[{\rm mas}]$  & $[{\rm mas}]$ & $[{\rm Jy\>beam^{-1}}]$ & $[{\rm km\>s^{-1}}]$ & \\ 
\hline
1 & 1a & $0.35\,\pm\,0.06$ & $-0.17\,\pm\,0.08$ & $0.836$--$2.698$ & $-48.7$ & 123456789**\\
& 1b & $0.41\,\pm\,0.05$ & $-0.28\,\pm\,0.08$ & $1.278$--$5.461$ & $-49.1$ & 123456789**\\
& 1c & $0.48\,\pm\,0.08$ & $-0.33\,\pm\,0.14$ & $1.569$--$8.010$ & $-49.5$ & 123456*****\\
\hline
2 & 2a & $0.54\,\pm\,0.03$ & $-0.44\,\pm\,0.10$ & $3.633$--$9.049$ & $-49.9$ & 12345******\\
& 2b & $0.58\,\pm\,0.03$ & $-0.47\,\pm\,0.09$ & $1.088$--$3.577$ & $-50.4$ & 12345******\\
\hline
3 & 3a & $3.47\,\pm\,0.08$ & $-2.38\,\pm\,0.11$ & $2.376$--$3.649$ & $-60.9$ & 12345******\\
 & 3b & $3.58\,\pm\,0.08$ & $-2.35\,\pm\,0.11$ & $3.455$--$16.154$ & $-61.3$ & 123456*****\\
\hline
4 & 4a & $4.42\,\pm\,0.09$ & $-2.29\,\pm\,0.16$ & $1.474$--$24.991$ & $-61.7$ & ***456789**\\
& 4b & $4.47\,\pm\,0.09$ & $-2.28\,\pm\,0.14$ & $1.328$--$27.269$ & $-62.2$ & ***456789**\\
\hline
5 & 5a & $2.57\,\pm\,0.05$ & $-2.41\,\pm\,0.18$ & $0.741$--$4.443$ & $-56.7$ & ***45678***\\
& 5b & $2.51\,\pm\,0.07$ & $-2.31\,\pm\,0.31$ & $1.349$--$2.558$ & $-57.1$ & ****5678***\\
\hline
6 & 6a & $8.20\,\pm\,0.13$ & $3.68\,\pm\,0.17$ & $0.662$--$11.723$ & $-59.6$ & *****6789AB\\
& 6b & $8.18\,\pm\,0.17$ & $3.76\,\pm\,0.13$ & $1.444$--$25.803$ & $-60.1$ & *****6789AB\\
& 6c & $8.20\,\pm\,0.11$ & $3.78\,\pm\,0.14$ & $8.174$--$33.455$ & $-60.5$ & *****6789AB\\
& 6d & $8.18\,\pm\,0.11$ & $3.71\,\pm\,0.18$ & $3.146$--$33.106$ & $-60.9$ & *****6789AB\\
       \hline
     \hline
    \end{tabular}
  \end{center}
  {$^{\ast}$$(\alpha_0,\delta_0)$ shows the position of the maser feature on DOY 1 since 2012. Digits and letters in the last column of "Detection" denote epochs of detection, which can be identified with table \ref{sum_obs}.}
\end{table*}

\begin{table*}
  \caption{Parallax results for features 1, 4 and 6$^{\ast}$}\label{feat_146}
  \begin{center}
    \begin{tabular}{lcrrrl}
      \hline
      \hline
 DOY  & Spot  & Parallax & $\chi^2$ & $V_{\rm LSR}$ & Epochs  \\
$[{\rm days}]$ &  & $[{\rm mas}]$ &  & $[{\rm km\>s^{-1}}]$ &  \\
      \hline
87--874  & 1a & $0.146\,\pm\,0.040$ & $1.410$ & $-48.7$ & 123456789**\\
 & 1b & $0.143\,\pm\,0.039$ & $1.379$ & $-49.1$ & 123456789**\\
 & 1c & $0.152\,\pm\,0.062$ & $2.401$ & $-49.5$ & 123456*****\\
 \hline
\multicolumn{2}{c}{feature 1 combined} & $0.146\,\pm\,0.044$ & $1.517$ & & \\ 
 \hline
390--874 & 4a & $0.142\,\pm\,0.044$ & $0.927$ & $-61.7$ & ***456789**\\
 & 4b & $0.155\,\pm\,0.041$ & $0.823$ & $-62.2$ & ***456789**\\
\hline
\multicolumn{2}{c}{feature 4 combined} & $0.148\,\pm\,0.041$ & $0.820$ & &\\ 
\hline
592--1044 & 6a & $0.150\,\pm\,0.041$ & $0.799$ & $-59.6$ & *****6789AB\\
 & 6b & $0.141\,\pm\,0.045$ & $0.954$ & $-60.1$ & *****6789AB\\
 & 6c & $0.164\,\pm\,0.034$ & $0.534$ & $-60.5$ & *****6789AB\\
 & 6d & $0.181\,\pm\,0.036$ & $0.612$ & $-60.9$ & *****6789AB\\
 \hline
\multicolumn{2}{c}{feature 6 combined} & $0.159\,\pm\,0.038$ & $0.669$ & \\ 
 \hline
  \hline
 \multicolumn{2}{c}{All spots combined} & $0.151\,\pm\,0.042$ & $1.039$ & \\
       \hline
     \hline
    \end{tabular}
  \end{center}
  {$^\ast$DOY (day of year), parallax, and LSR velocity $V_{\rm LSR}$, of spots 1a--1c, 4a--4b and 6a--6d. The combined parallax was obtained using the spots 1a--1c, 4a--4b and 6a--6d. We applied the errors of R.A. ($0.073$) and Decl. (0.107) to each spot and feature. Parallaxes of each spot are consistent within the range of error.} For the error of parallax, we multiplied it by a square root of the number of maser spots $\sqrt{N_{\rm spot}}$ since the maser spot data are highly correlated, caused by similar atmospheric delay differences between maser spots and a background QSO (see figure \ref{rafit_decfit}).
\end{table*}

\begin{table*}
  \caption{Determination of the systematic proper motion for IRAS$\,$01123+6430$^{\ast}$}\label{feat_123456}
  \begin{center}
    \begin{tabular}{lcrrr}
      \hline
      \hline
  Feature  & DOY & $\mu_\alpha{\rm cos}\delta$ & $\mu_\delta$ & $V_{\rm LSR}$  \\
  & $[{\rm days}]$ & $[{\rm mas\>yr^{-1}}]$ & $[{\rm mas\>yr^{-1}}]$ & $[{\rm km\>s^{-1}}]$ \\
      \hline
  1 & 87--874 & $-1.57\,\pm\,0.03$ & $-0.20\,\pm\,0.04$ & $-49.5$--$-48.7$\\
  2 & 87--434 & $-1.74\,\pm\,0.03$ & $0.10\,\pm\,0.08$ & $-50.4$--$-49.9$\\  
 3 & 87--592 & $-1.03\,\pm\,0.06$ & $-0.66\,\pm\,0.08$ & $-61.3$--$-60.9$\\ 
 4 & 390--874 & $-1.70\,\pm\,0.03$ & $-0.72\,\pm\,0.05$ & $-62.2$--$-61.7$\\ 
 5 & 390--812 & $-1.76\,\pm\,0.02$ & $-0.48\,\pm\,0.08$ & $-57.1$--$-56.7$\\ 
 6 & 592--1044 & $-0.82\,\pm\,0.03$ & $0.35\,\pm\,0.03$ & $-60.9$--$-59.6$\\
  \hline
 Average &  & $-1.44\,\pm\,0.15$ & $-0.27\,\pm\,0.16$ & $-57.1$\\
       \hline
     \hline
    \end{tabular}
  \end{center}
  {$^\ast$DOY, Proper motion ($\mu_\alpha{\rm cos}\delta$, $\mu_\delta$) and LSR velocity $V_{\rm LSR}$, of features 1, 2, 3, 4, 5 and 6. Using these features, we calculated the average of the proper motions. Errors in the averaged proper motion are calculated by the standard deviation of proper motions.}
\end{table*}

\subsection{Parallax and proper motion}

\begin{figure*}
\begin{center}
\FigureFile(55mm,55mm){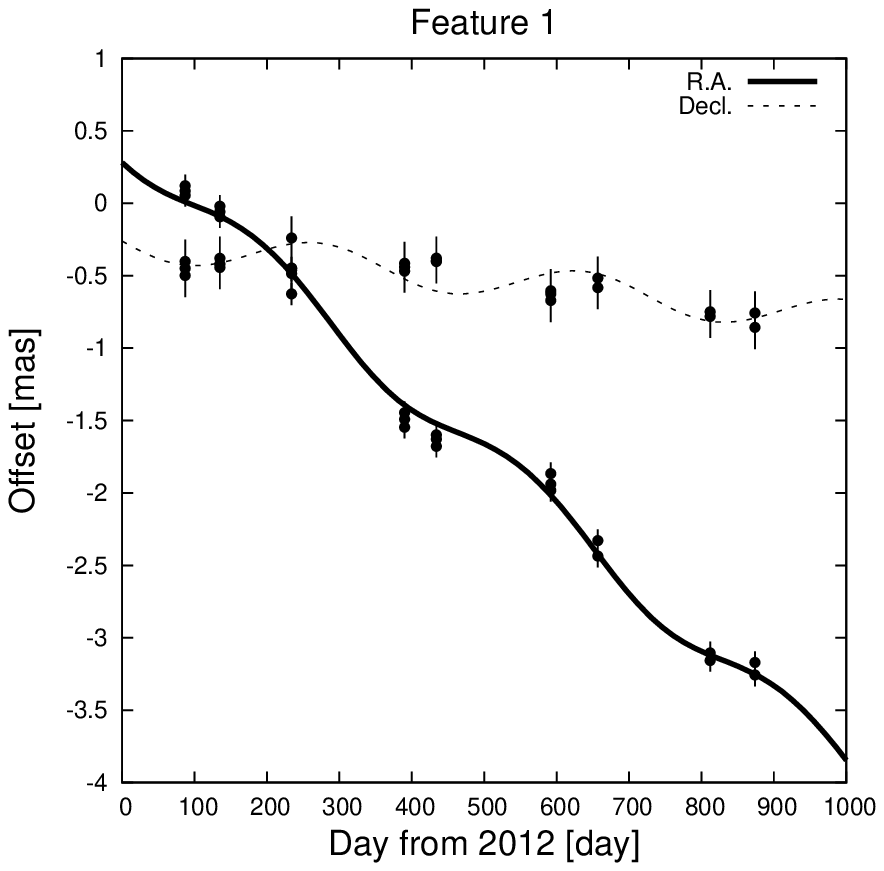}
\FigureFile(55mm,55mm){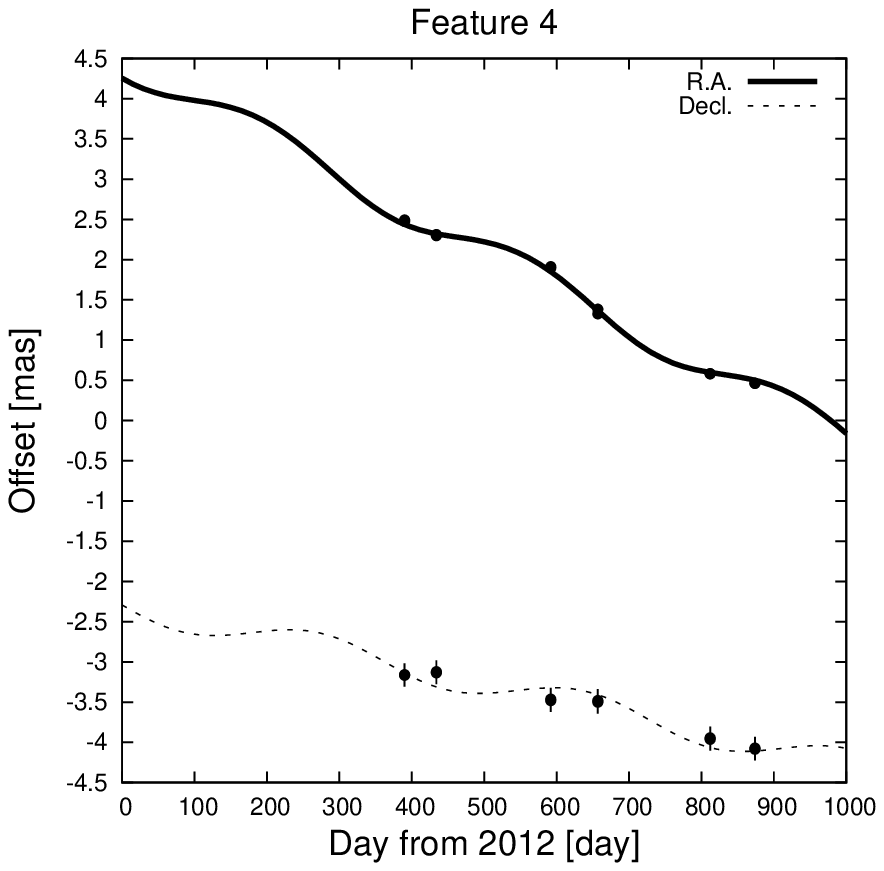}
\FigureFile(55mm,55mm){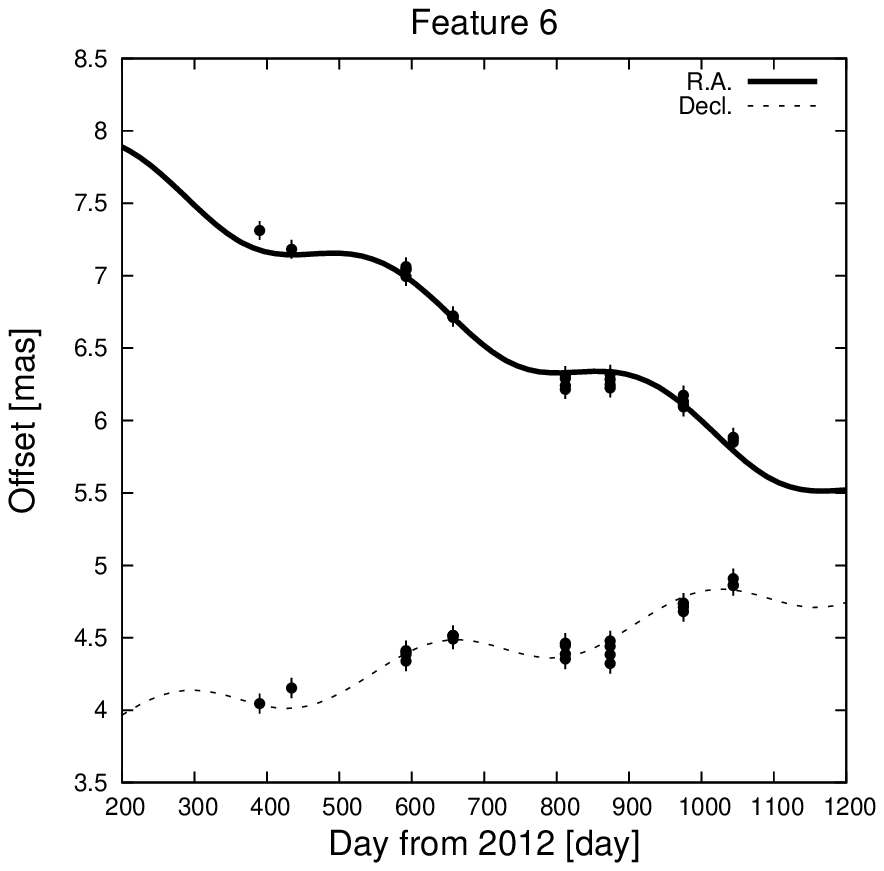}
\end{center}
\caption{Observed motions (proper motion + parallax) of IRAS$\,$01123+6430 as a function of time for feature 1 (left), feature 4 (middle) and feature 6 (right). The full and dashed lines show the motions in right ascension and declination, respectively.}
\label{rafit_decfit} 
\end{figure*}

\begin{figure}[htbp]
\begin{center}
\includegraphics[width=9.0cm,clip]{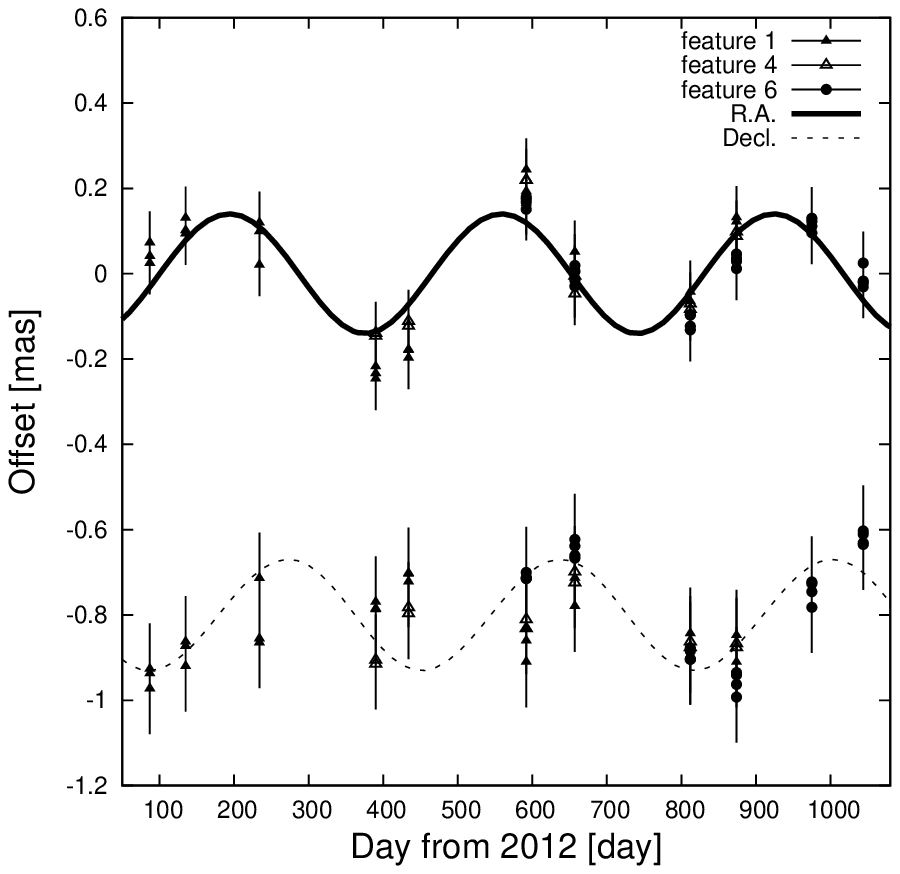}
\end{center}
\caption{Time variation of offsets in right ascension and declination after subtracting the proper motions from the observed position offsets.}
\label{rafit_decfit2} 
\end{figure}

\begin{figure*}
\begin{center}
\FigureFile(170mm,170mm){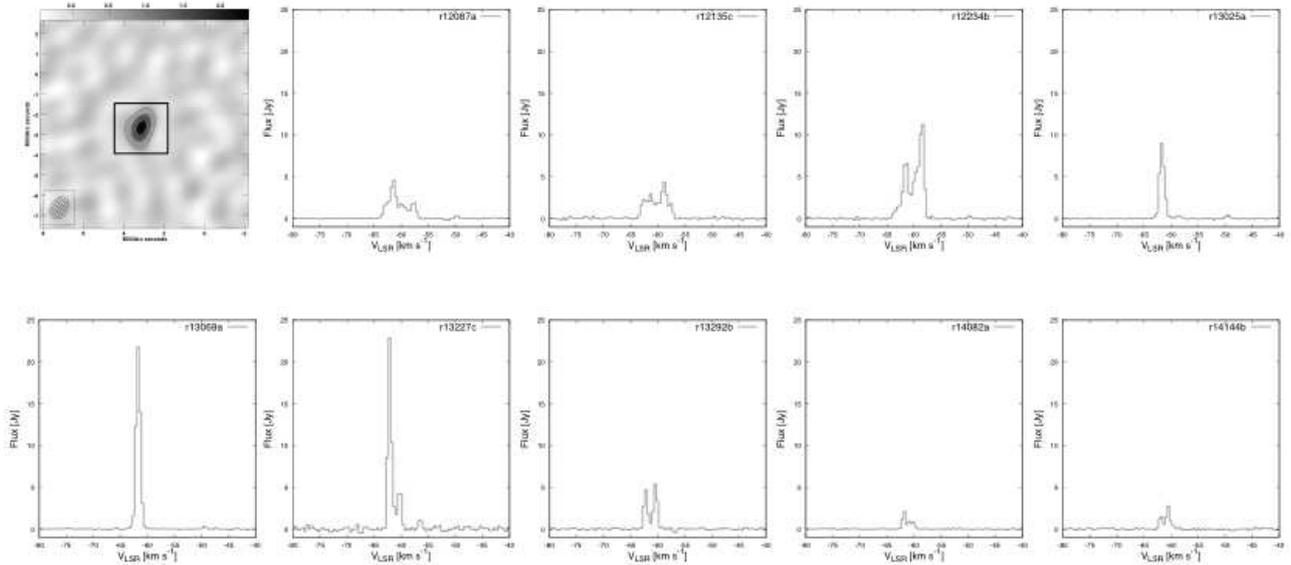}
\end{center}
\caption{A contour map of the feature 4 (integrated from $V_{\rm{LSR}}\ =\ -62.2$ to $-61.7\>{\rm km\>s^{-1}}$) for r12135c epoch (top left) and spectra with CLEANed image cube of the feature 4 from r12087a to r14144b (the others). The spectra show broad line width of $\sim 6$--$7\>{\rm km\>s^{-1}}$ contributed from feature 3 for the first 3 epochs, while the line width narrowed to $\sim 2$--$3\>{\rm km\>s^{-1}}$ in the later six epochs.}
\label{spots_image_feat4}
\end{figure*}

\begin{figure*}
\begin{center}
\FigureFile(55mm,55mm){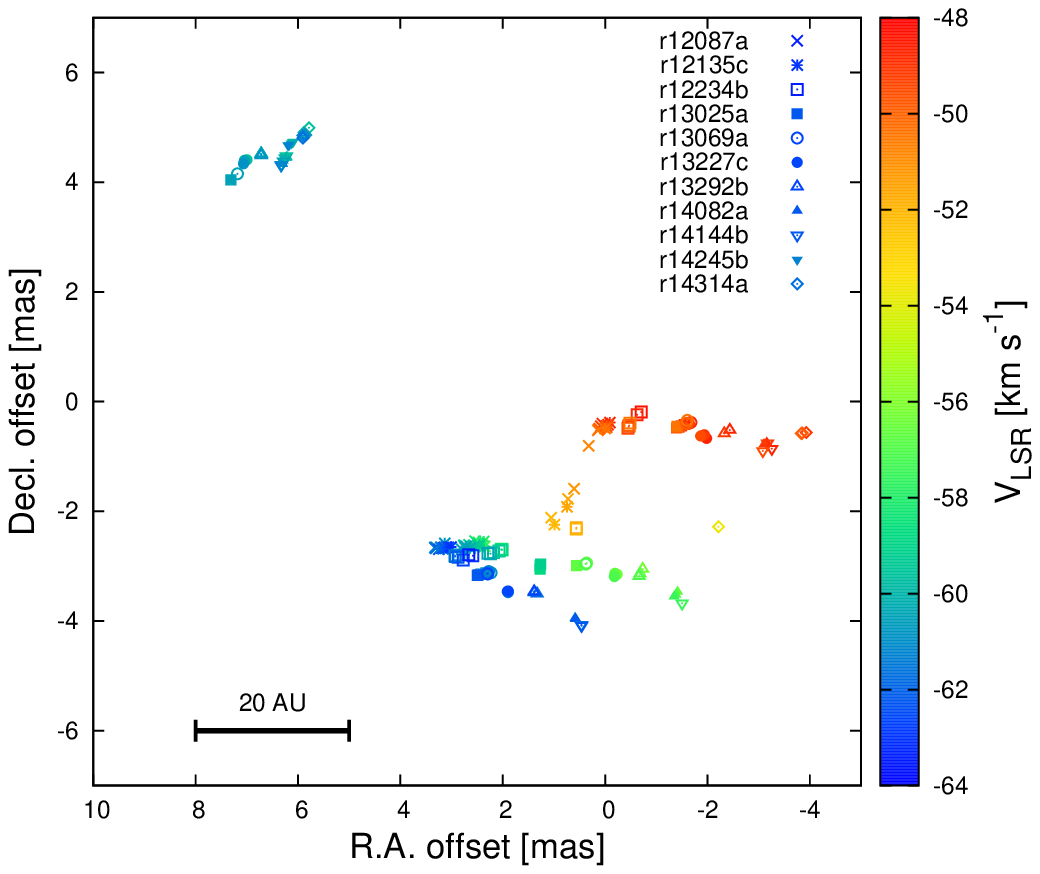}
\FigureFile(55mm,55mm){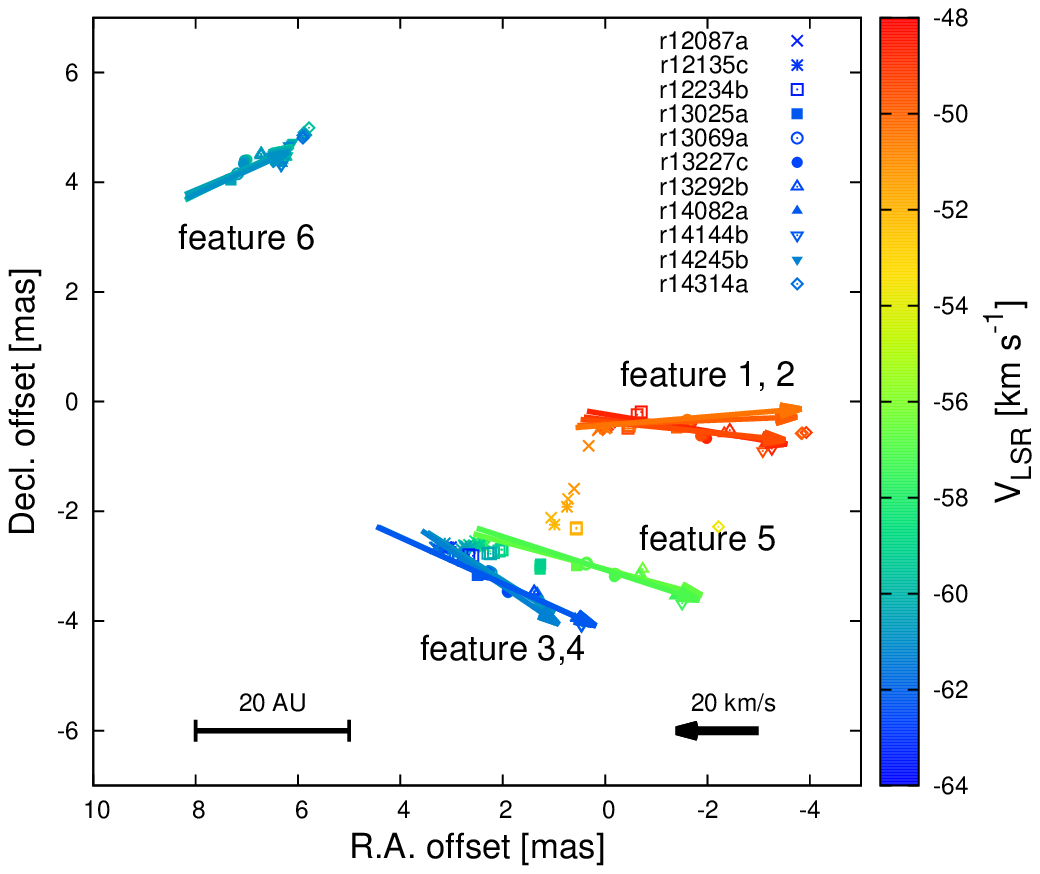}
\FigureFile(55mm,55mm){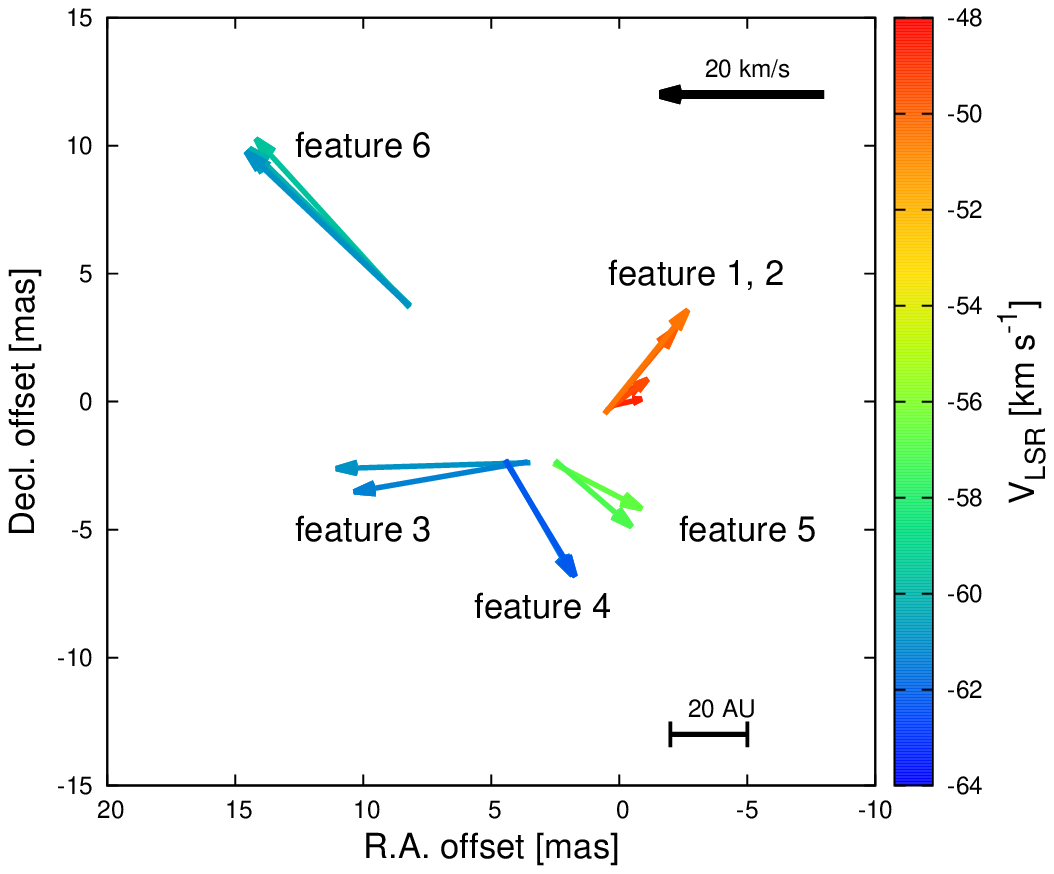}
\end{center}
\caption{(Left) The distribution map of all detected maser spots (the kinds of points show individual epochs "${\rm r}\ast\ast\ast\ast\ast\ast$"), the color bar shows LSR velocity, and the black bar shows 20$\>$AU scale at a distance of 6.61$\>$kpc. (Middle) The distribution map of all detected maser spots with the vectors of each feature's proper motion showed by arrows (black arrow shows $20\>{\rm km\>s^{-1}}$ scale). (Right) The internal motion map obtained by subtracting the systematic proper motions from the original proper motion map.}\label{maser_dist_vect}
\end{figure*}

We plot position offsets of maser features 1, 4 and 6 as a function of time in figure \ref{rafit_decfit}, where fitted model lines are also plotted. The resultant curve of the annual parallax obtained by combined fitting of all the features is shown in figure \ref{rafit_decfit2}, where time variations of offsets in right ascension and declination after subtracting the proper motions are plotted. The combined parallax was obtained to be $\varpi \ =\ 0.151\,\pm\,0.042\>{\rm mas}$, which corresponds to a trigonometric distance of $D\ =\ 6.61^{+2.55}_{-1.44}\>{\rm kpc}$. Measured parallaxes for all spots and combined fitting are consistent with each other within the error. It should be noted that we did not include data for feature 4 in the time range of DOY=87--234 because spectra of the feature 4 showed broad lines in this time range as shown in figure \ref{spots_image_feat4} while it showed narrow line after DOY=234. Such a broad line seems to indicate that the spots are extended or blended with other spots. Therefore, we concluded that the parallax for feature 4 should be measured using only the data of DOY=390--874, otherwise we found that inconsistent parallax is derived. The distribution of maser spots and vectors of individual proper motions are shown in the left panel of figure \ref{maser_dist_vect}.

We also estimated the mean proper motion of this system by averaging the proper motions of the features 1--6 (table \ref{feat_123456}) followed by a linear fit to the astrometric data. Hereafter we call this the systematic proper motion. The right panel of figure \ref{maser_dist_vect} shows internal motion vectors of individual maser features, which were obtained by subtracting the systematic proper motion from motion vectors shown in the middle panel of figure \ref{maser_dist_vect}. The values of the internal proper motions are listed in table \ref{inter_motion}.

The resultant systematic proper motion components in equatorial coordinates were $(\mu_\alpha{\rm cos}\delta,\ \mu_\delta)\ =\ (-1.44\,\pm\,0.15,\ -0.27\,\pm\,0.16)\>{\rm mas\>yr^{-1}}$, which corresponds to $(\mu_l{\rm cos}b,\ \mu_b)\ =\ (-1.40\,\pm\,0.14,\ -0.40\,\pm\,0.17)\>{\rm mas\>yr^{-1}}$ in Galactic coordinates. We converted the systematic proper motion (${\rm mas\>yr^{-1}}$) into a linear velocity scale (${\rm km\>s^{-1}}$) based on the resultant heliocentric distance of 6.61$\>$kpc. It corresponded to $v_l\ =\ D\mu_l{\rm cos}b\ =\ -44.0\,\pm\,17.5\>{\rm km\ s^{-1}}$ and $v_b\ =\ D\mu_b\ =\ -12.6\,\pm\,7.3\>{\rm km\>s^{-1}}$.

\begin{table*}
  \caption{Internal proper motions of maser features in IRAS$\,$01123+6430$^{\ast}$}\label{inter_motion}
  \begin{center}
    \begin{tabular}{llrrrrrr}
      \hline
      \hline
Feature & ID & \multicolumn{2}{c}{R.A.} & & \multicolumn{2}{c}{Decl.} & $V_{\rm LSR}$\\
\cline{3-4} \cline{6-7}
 &  & $\mu_\alpha{\rm cos}\delta-\langle \mu_\alpha{\rm cos}\delta\rangle$ & $v_\alpha-\langle v_\alpha\rangle$ & & $\mu_\delta-\langle \mu_\delta\rangle$ & $v_\delta-\langle v_\delta\rangle$ & \\
 &  & $[{\rm mas\>yr^{-1}}]$ & $[{\rm km\>s^{-1}}]$ & & $[{\rm mas\>yr^{-1}}]$ & $[{\rm km\>s^{-1}}]$ & $[{\rm km\>s^{-1}}]$\\ 
\hline
1 & 1a  & $-0.12\,\pm\,0.16$ & $-3.8\,\pm\,4.9$ & & $0.03\,\pm\,0.17$ & $0.9\,\pm\,5.3$ & $-48.7$\\
& 1b  & $-0.12\,\pm\,0.16$ & $-3.6\,\pm\,4.9$ & & $0.09\,\pm\,0.17$ & $2.8\,\pm\,5.3$ & $-49.1$\\
& 1c  & $-0.16\,\pm\,0.17$ & $-5.0\,\pm\,5.4$ &  & $0.12\,\pm\,0.21$ & $3.8\,\pm\,6.6$ & $-49.5$\\
\hline
2 & 2a &  $-0.28\,\pm\,0.16$ & $-8.6\,\pm\,4.9$ & & $0.33\,\pm\,0.21$ & $10.3\,\pm\,6.5$ & $-49.9$\\
& 2b & $-0.32\,\pm\,0.16$ & $-10.2\,\pm\,4.9$ & & $0.40\,\pm\,0.20$ & $12.7\,\pm\,6.1$ & $-50.4$\\
\hline
3 & 3a &  $0.76\,\pm\,0.18$ & $23.7\,\pm\,5.7$ & & $-0.02\,\pm\,0.21$ & $-0.7\,\pm\,6.6$ & $-60.9$\\
 & 3b &  $0.68\,\pm\,0.17$ & $21.2\,\pm\,5.3$ & & $-0.12\,\pm\,0.19$ & $-3.7\,\pm\,6.0$ & $-61.3$\\
 \hline
4 & 4a &  $-0.26\,\pm\,0.16$ & $-8.1\,\pm\,5.0$ & & $-0.45\,\pm\,0.18$ & $-14.1\,\pm\,5.7$ & $-61.7$\\
& 4b &  $-0.27\,\pm\,0.16$ & $-8.6\,\pm\,5.0$ & & $-0.45\,\pm\,0.18$ & $-14.2\,\pm\,5.5$ & $-62.2$\\
\hline
5 & 5a &  $-0.34\,\pm\,0.16$ & $-10.8\,\pm\,4.9$ & & $-0.18\,\pm\,0.19$ & $-5.5\,\pm\,6.0$ & $-56.7$\\
& 5b & $-0.30\,\pm\,0.16$ & $-9.3\,\pm\,4.9$ & & $-0.26\,\pm\,0.24$ & $-8.1\,\pm\,7.5$ & $-57.1$\\
\hline
6 & 6a &  $0.60\,\pm\,0.16$	 & $18.7\,\pm\,5.1$ & & $0.66\,\pm\,0.18$ & $20.6\,\pm\,5.5$ & $-59.6$\\
& 6b & $0.63\,\pm\,0.17$ & $19.6\,\pm\,5.3$ & & $0.61\,\pm\,0.17$ & $19.2\,\pm\,5.3$ & $-60.1$\\
& 6c &  $0.62\,\pm\,0.16$ & $19.4\,\pm\,5.0$ & &  $0.59\,\pm\,0.17$ & $18.4\,\pm\,5.3$ & $-60.5$\\
& 6d &  $0.64\,\pm\,0.16$ & $20.0\,\pm\,5.0$ & & $0.61\,\pm\,0.18$ & $18.9\,\pm\,5.6$ & $-60.9$\\
       \hline
     \hline
    \end{tabular}
  \end{center}
  {$^{\ast}$ $(\mu_\alpha{\rm cos}\delta,\ \mu_\delta)$ shows the individual feature's proper motions, and $(\langle \mu_\alpha{\rm cos}\delta\rangle,\ \langle \mu_\delta\rangle)$ shows the systematic proper motion.}
\end{table*}

\subsection{The Galactocentric distance and rotation velocity}

Using the Galactocentric distance of the Sun, $R_0\ =\ 8.05\,\pm\,0.45\>{\rm kpc}$ \citep{2012PASJ...64..136H}, Galactic coordinates $(l,\ b)\ =\ (\timeform{125.51D},\ \timeform{+2.03D})$, and the trigonometric distance $D\ =\ 6.61^{+2.55}_{-1.44}\>{\rm kpc}$, the Galactocentric distance of IRAS$\,$01123+6430 was calculated to be $R=13.04\,\pm\,2.24\>{\rm kpc}$. The LSR velocity $-54.7\>{\rm km\>s^{-1}}$ \citep{1989A&AS...80..149W} transformed into heliocentric velocity is $V_{\rm{Helio}}\ =\ -61.4\>{\rm km\>s^{-1}}$, based on the standard solar motion (see below). Next, we calculated the three dimensional velocity vectors $(U,\ V,\ W)$ of IRAS$\,$01123+6430 by adopting the Galactic constants $(R_0,\ \Theta_0)\ =\ (8.05\,\pm\,0.45\>{\rm kpc},\ 238\,\pm\,14\>{\rm km\>s^{-1}})$ \citep{2012PASJ...64..136H}, and the standard solar motion $(U^{\rm Std}_{\odot},V^{\rm Std}_{\odot},W^{\rm Std}_{\odot})=(10.3,15.3,7.7)$ $\rm{km\ s^{-1}}$ \citep{1986MNRAS.221.1023K}. From these values, $V_{\rm{Helio}}$, $v_l$, and $v_b$, we derived the 3-D velocity vectors $(U,\ V,\ W)\ =\ (81.2\,\pm\,14.2,\ 226\,\pm\,17,\ -7.5\,\pm\,7.3)\>{\rm km\>s^{-1}}$, which corresponds to $(V_{R_{\rm in}},\ V_{\theta},\ V_z)\ =\ (-19.1\,\pm\,5.8,\ 239\,\pm\,22,\ -7.5\,\pm\,7.3)\>{\rm km\>s^{-1}}$ in cylindrical coordinates. A detailed explanation of coordinates and transformation of the 3-D velocity vectors is given in \citet{2015PASJ...67...68N}. Regarding the calculation of errors, see the Appendix.

As a result, the Galactocentric distance, $R$, and the rotation velocity, $\Theta$, of IRAS$\,$01123+6430 were determined as $(R,\ \Theta)\ =\ (13.04\,\pm\,2.24\>{\rm kpc},\ 239\,\pm\,22\>{\rm km\>s^{-1}})$. For comparison, we also calculated $R$, $\Theta$ with other Galactic constants taken from \citet{2014ApJ...783..130R} and \citet{1986MNRAS.221.1023K}, which are summarized in table \ref{sum_const_3dmotn}. The value of $V_{\theta}\ -\ \Theta_0$ is within the range of error, which is consistent with a flat rotation curve.

\begin{table*}
  \caption{Galactocentric distance, $R$, and 3-D velocity in the cylindrical coordinates, $(V_{R_{\rm in}},\ V_{\theta},\ V_z)$, and some possible Galactic constants, $(R_0,\ \Theta_0)$, from each reference$^{\ast}$}\label{sum_const_3dmotn}
  \begin{center}
    \begin{tabular}{llllllrl}
      \hline
      \hline
  $R_0$  & $\Theta_0$ & $R$ & $V_{R_{\rm in}}$ & $V_{\theta}$  & $V_z$ & $V_{\rm \theta}\ -\ \Theta_0$ & Reference \\
 $[{\rm kpc}]$ & $[{\rm km\>s^{-1}}]$ & $[{\rm kpc}]$ & $[{\rm km\>s^{-1}}]$ & $[{\rm km\>s^{-1}}]$ & $[{\rm km\>s^{-1}}]$ & $[{\rm km\>s^{-1}}]$ &  \\
      \hline
 $8.05\,\pm\,0.45$ & $238\,\pm\,14$ & $13.04\,\pm\,2.24$ & $-19.1\,\pm\,5.8$ & $239\,\pm\,22$ & $-7.5\,\pm\,7.3$ & $1$ & \citet{2012PASJ...64..136H}\\
 8.5 & 220 & $13.5\,\pm\,2.2$ & $-8.6\,\pm\,9.0$ & $223\,\pm\,15$ & $-7.5\,\pm\,7.3$ & $3$ & \citet{1986MNRAS.221.1023K}\\ 
 $8.34\,\pm\,0.16$ & $240\,\pm\,8$ & $13.31\,\pm\,2.19$ & $-17.8\,\pm\,7.8$ & $241\,\pm\,18$ & $-7.5\,\pm\,7.3$ & $1$ & \citet{2014ApJ...783..130R}\\
       \hline
     \hline
    \end{tabular}
  \end{center}
  {$^{\ast}$We used the Galactic constants referred to \citet{2012PASJ...64..136H}, \citet{1986MNRAS.221.1023K}, and \citet{2014ApJ...783..130R}.}
\end{table*}

\section{Discussion}

\subsection{Spectral type of the young stellar object}

We estimated the bolometric luminosity of the young stellar object (YSO) using the newly obtained trigonometric distance and mid- and far-infrared flux densities. According to \citet{1988SSSC..C......0H}, the infrared flux densities of IRAS$\,$01123+6430 are $f_{\rm 12\mu m}\ =\ 8.34\ \times\ 10^{-1}\>{\rm Jy}$, $f_{\rm 25\mu m}\ =\ 1.49\>{\rm Jy}$, $f_{\rm 60\mu m}\ =\ 18.8\>{\rm Jy}$, and $f_{\rm 100\mu m}\ =\ 65.0\>{\rm Jy}$. Based on the formula, $L_{\rm Bol}\ =\ 5.4D^2\ (f_{12\mu{\rm m}}\ /\ 0.79\ +\ f_{25\mu{\rm m}}\ /\ 2\ +\ f_{60\mu{\rm m}}\ /\ 3.9\ +\ f_{100\mu{\rm m}}\ /\ 9.9)\>L_{\odot}$, presented by \citet{2009A&A...507..369W}, the bolometric luminosity of the YSO is calculated to be $L_{\rm Bol}\ =\ (3.11\,\pm\,2.86)\ \times\ 10^3\>L_{\odot}$, which corresponds to the spectral type of B1--B2 \citep{1973AJ.....78..929P}. The spectral type of the source is earlier than a B3 star, which is consistent with the development of an H$\,$\emissiontype{II} region in IRAS$\,$01123+6430.

\subsection{Internal motion}

The internal motions are presented as a vector map shown in the right panel of figure \ref{maser_dist_vect}, which shows that all features move outward from $(\Delta \alpha,\ \Delta \delta)\ \sim\ (3,\ 0)\>{\rm mas}$. The LSR velocity of features 3 and 4 are blue-shifted and those of features 1 and 2 are red-shifted against the systemic velocity of $-54.7\>{\rm km\>s^{-1}}$ \citep{1989A&AS...80..149W}. This internal motion resembles a bipolar outflow from the central YSO, which is commonly presented in former works (e.g., \cite{2016PASJ...68...69M}; \cite{2015PASJ...67...68N}). Meanwhile, the extent of the maser distribution shown in figure \ref{maser_dist_vect} is about 50$\>$AU whereas the physical sizes of most bipolar outflows are typically more than 100$\>$AU (e.g. \cite{2016MNRAS.460..283B}; \cite{2005PASJ...57..595H}; \cite{2018ApJ...865L..12B}; \cite{2018MNRAS.tmp.2852B}; \cite{2018ApJ...865...37M}; \cite{2018AJ....156..239Z}; \cite{2018RAA....18...93Z}). Therefore, internal motions of features 1--6 could either represent a very young, compact ejection resembling the case of \citet{2018ApJ...863...94L}, or an expanding circumstellar disk. 

\subsection{The position of IRAS$\,$01123+6430 in the Galaxy based on the trigonometric distance}

If we calculate the probability distance of the IRAS$\,$01123+6430 based on the LSR velocity of $V_{\rm LSR}\ =\ -54.7\>{\rm km\>s^{-1}}$ using the Bayesian estimator\footnotemark \citep{2016ApJ...823...77R}, we obtain a distance of 4.51$\>{\rm kpc}$, which corresponds to distance of the inter-arm region between the Perseus and Cygnus arm (Outer arm) \citep{2016PASJ...68....5N}. On the other hand, the annual parallax indicates that the source is located around the Cygnus arm (Outer arm) further than the inter-arm region. The discrepancy is caused by the radial motion of $V_{R_{\rm in}}\ =\ -19.1\,\pm\,5.8\>{\rm km\>s^{-1}}$, which reduces the absolute value of $V_{\rm LSR}$. If we considered that IRAS$\,$01123+6430 would follow pure circular motion around the Galaxy with a rotational velocity of $V_\theta\ =\ 239\>{\rm km\>s^{-1}}$ omitting the radial velocity, the LSR velocity would be calculated as $V_{\rm LSR}\ =\ [(R_0/R)V_\theta-\Theta_0]{\rm sin}l{\rm cos}b=\ -73.6\>{\rm km\>s^{-1}}$, in assuming $R_0\ =\ 8.05\>{\rm kpc}$, $R\ =\ 13.04\>{\rm kpc}$, and $(l, b)=(\timeform{125.51D}, \timeform{+2.03D})$. When we calculated the probability distance based on the altered LSR velocity of $V_{\rm LSR}\ =\ -73.6\>{\rm km\>s^{-1}}$ using the same estimator, the probability distance is estimated to be 5.83$\>{\rm kpc}$, which corresponds to the location of the Outer arm. Therefore, our estimate is consistent with the formerly accepted kinematics and structures of the Galaxy.

\begin{figure*}
\begin{center}
\FigureFile(80mm,80mm){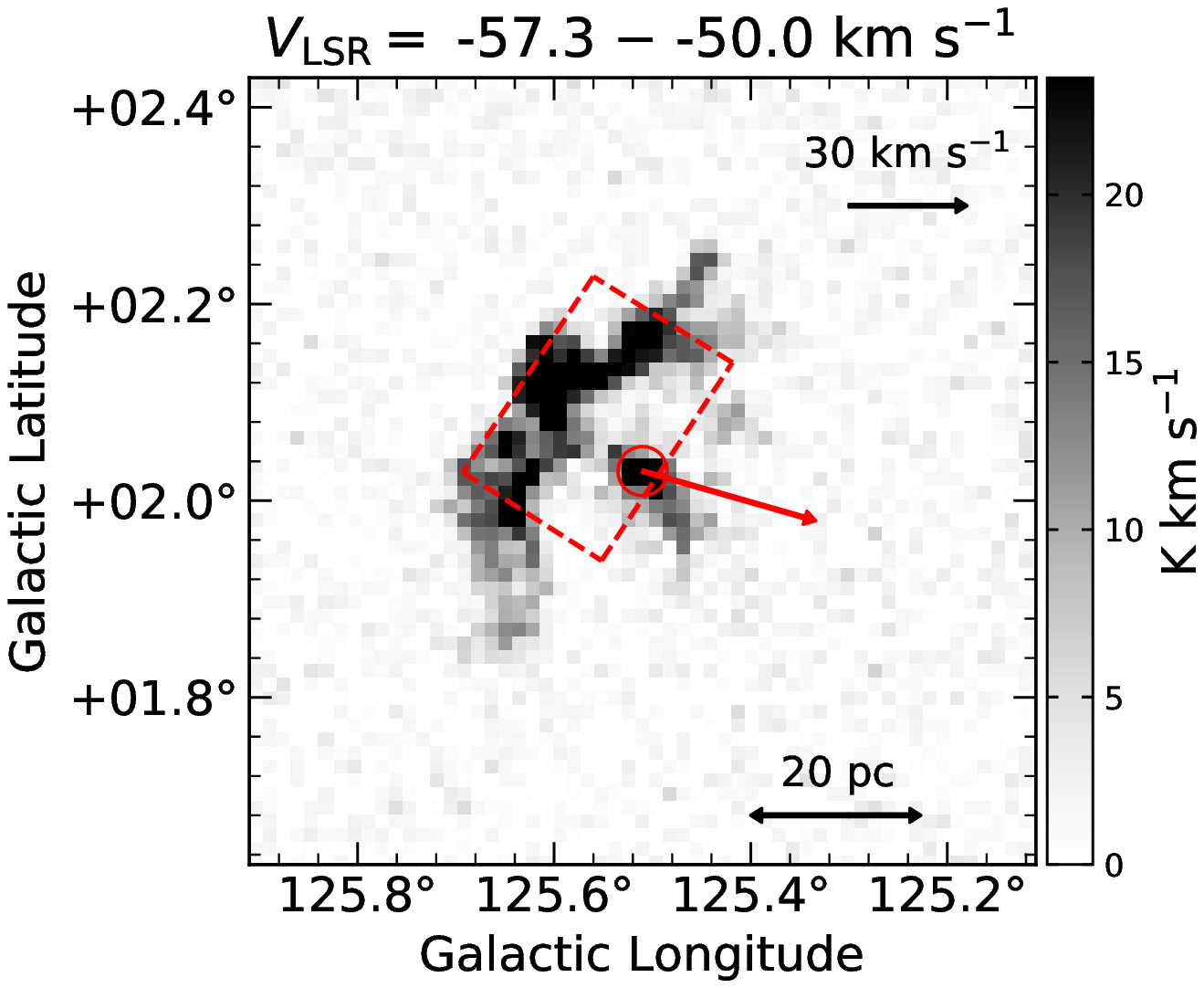}
\FigureFile(80mm,80mm){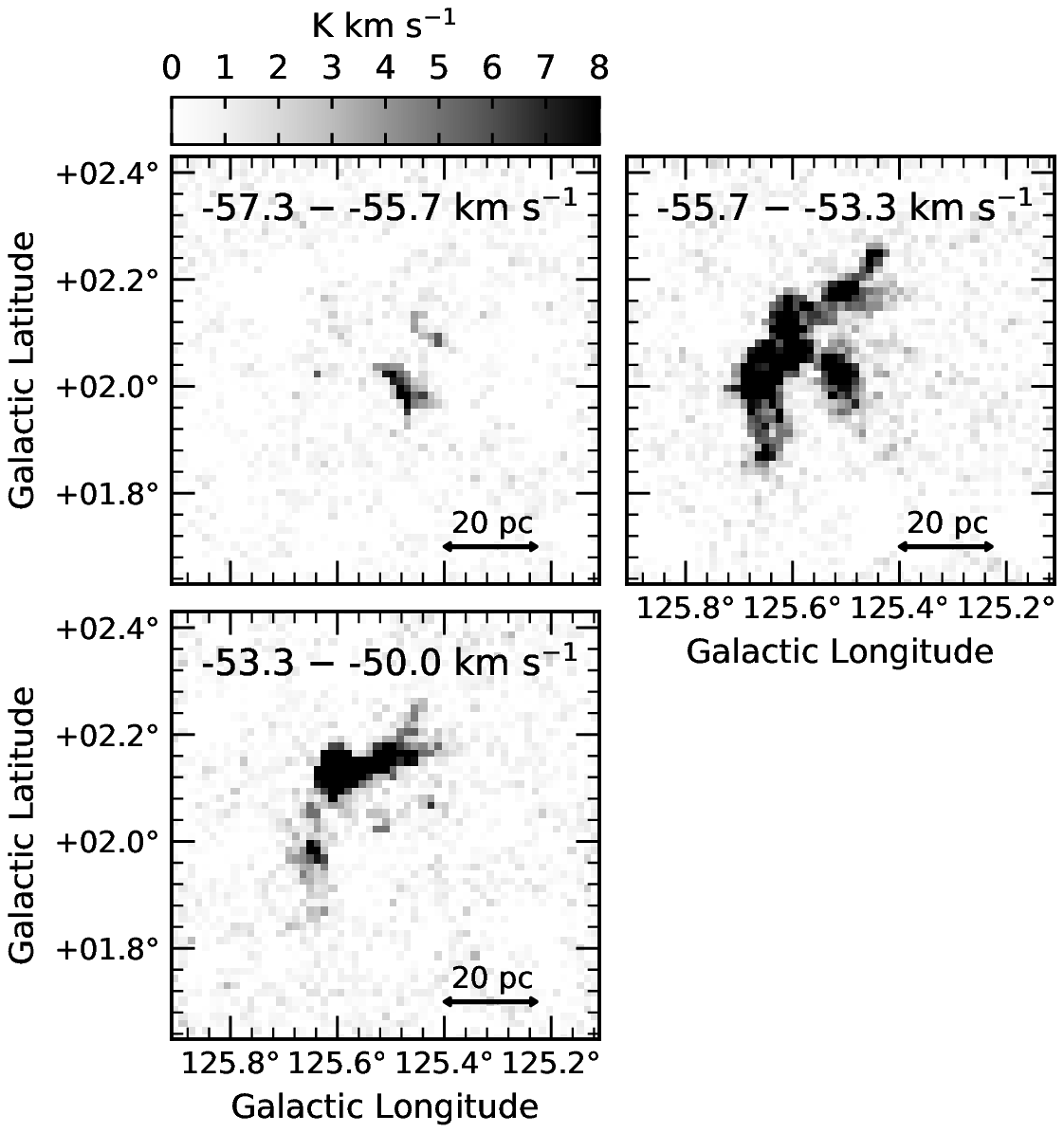}
\end{center}
\caption{(Left) Integrated map of HC$\,$5506 for $V_{\rm LSR}\ =\ -57.3$ to $-50.0\>{\rm km\>s^{-1}}$. The circle shows the position of IRAS$\,$01123+6430 (the grey bar shows the $\atom{CO}{}{12}$ $(J\,=\,1$--$0)$ integrated intensity), and the arrow shows the proper motion of IRAS$\,$01123+6430 in Galactic coordinates. (Right) The channel map of HC$\,$5506 observed by FCRAO \citep{1998ApJS..115..241H}, which is integrated at each velocity range from $-57.3$ to $-55.7\>{\rm km\>s^{-1}}$ (linear structure), $-55.7$ to $-53.3\>{\rm km\>s^{-1}}$ (medium structure), and $-53.3$ to $-50.0\>{\rm km\>s^{-1}}$ (arc-like structure).}\label{cloud_lb}
\end{figure*}

\footnotetext{http://bessel.vlbi-astrometry.org/bayesian}

\subsection{Associated molecular cloud: sign of cloud-cloud collision}

IRAS$\,$01123+6430 was observed in the $\atom{CO}{}{12}$ $(J\,=\,1$--$0)$ line with the FCRAO 14$\>$m telescope as a part of the FCRAO Outer Galaxy Survey \citep{1998ApJS..115..241H}, thus we can study the physical properties of the associated molecular cloud, which is catalogued as the 5506th cloud in the list. The molecular cloud was detected at the velocity range of $V_{\rm LSR}\ =\ -57.3$ to $-50.0\>{\rm km\>s^{-1}}$. Hereafter we refer to this cloud as HC$\,$5506 (Heyer's Cloud 5506). The CO line integrated intensity map of HC$\,$5506 is shown in the left panel of figure \ref{cloud_lb}.

\subsubsection{Physical parameters of HC$\,$5506}

The apparent size of this molecular cloud is $\Delta l\ \times\ \Delta b\ \sim\ \timeform{0.3D}\ \times\ \timeform{0.4D}$. Using the newly measured trigonometric distance of IRAS$\,$01123+6430, $D\ =\ 6.61^{+2.55}_{-1.44}\>{\rm kpc}$, the physical extent is calculated to be $35^{+13}_{-7}\>{\rm pc}\ \times\ 46^{+18}_{-10}\>{\rm pc}$. The CO luminosity of HC$\,$5506 was measured to be $L_{\rm CO}\ =\ (1.1\pm 1.0)\ \times\ 10^4\>{\rm K\>km\>s^{-1}\>pc^2}$ by integrating the CO emission over the region of antenna temperature $T^{\ast}_{R}\ \geq\ 0.5\>{\rm K}$ \citep{1998ApJS..115..241H}, and the cloud mass was calculated to be $M_{\rm{CO}}\ =\ (4.3\,\pm\,4.0)\ \times\ 10^4\>M_{\odot}$ using the conversion formula $M_{\rm CO}\ =\ 4.1\ ({L_{\rm CO}}\ /\ [{\rm K\>km\>s^{-1}\>pc^2}])$ \citep{2001ApJ...551..852H}.

HC$\,$5506 consists of two components, an arc-like and a linear structure, which can be clearly traced in the channel maps of $V_{\rm LSR}\ =\ -57.3$ to $-55.7\>{\rm km\>s^{-1}}$ and $V_{\rm LSR}\ =\ -53.3$ to $-50.0\>{\rm km\>s^{-1}}$ shown in the right panel of Figure \ref{cloud_lb}, respectively. IRAS$\,$01123+6430 is located at the north-eastern edge of the linear structure.

\begin{figure}[htbp]
\begin{center}
\includegraphics[width=8.0cm,clip]{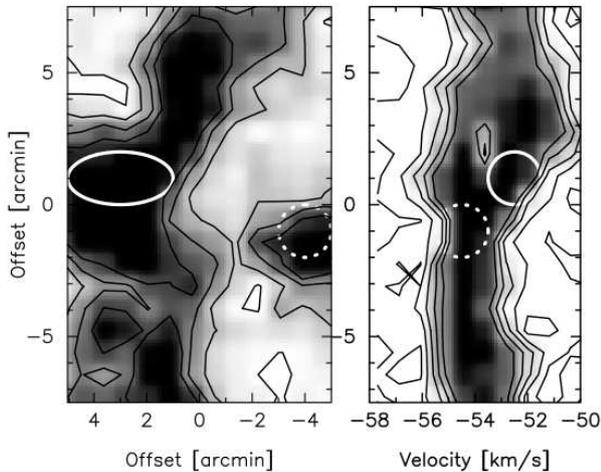}
\end{center}
\caption{(Left) The integrated map, cut out from the rectangle in the left panel of figure \ref{cloud_lb} and rotated $\timeform{35D}$ counter clockwise. (Right) The PV diagram integrated across the horizontal axis offsets on the left map. The white circle shows arc structure, and the dashed circle shows linear structure.}
\label{01123_V_B} 
\end{figure}

The left panel of figure \ref{cloud_lb} shows a region indicated with a rectangle to be extracted. In figure \ref{01123_V_B}, the left panel shows a zoomed-in view of the extracted region, and the right panel shows the position-velocity (PV) diagram obtained by integrating the extracted data cube in the south-west to the north-east direction. The PV diagram implies that the arc-like and linear structures differ in their central velocities by $2\>{\rm km\>s^{-1}}$ but this velocity separation is smaller than the line width of HC$\,$5506 system, $7\>{\rm km\>s^{-1}}$. The arc-like and linear structure are connected with each other in the PV diagram which implies that the these two structures are currently physically connected.

\begin{figure*}
\begin{center}
\FigureFile(75mm,75mm){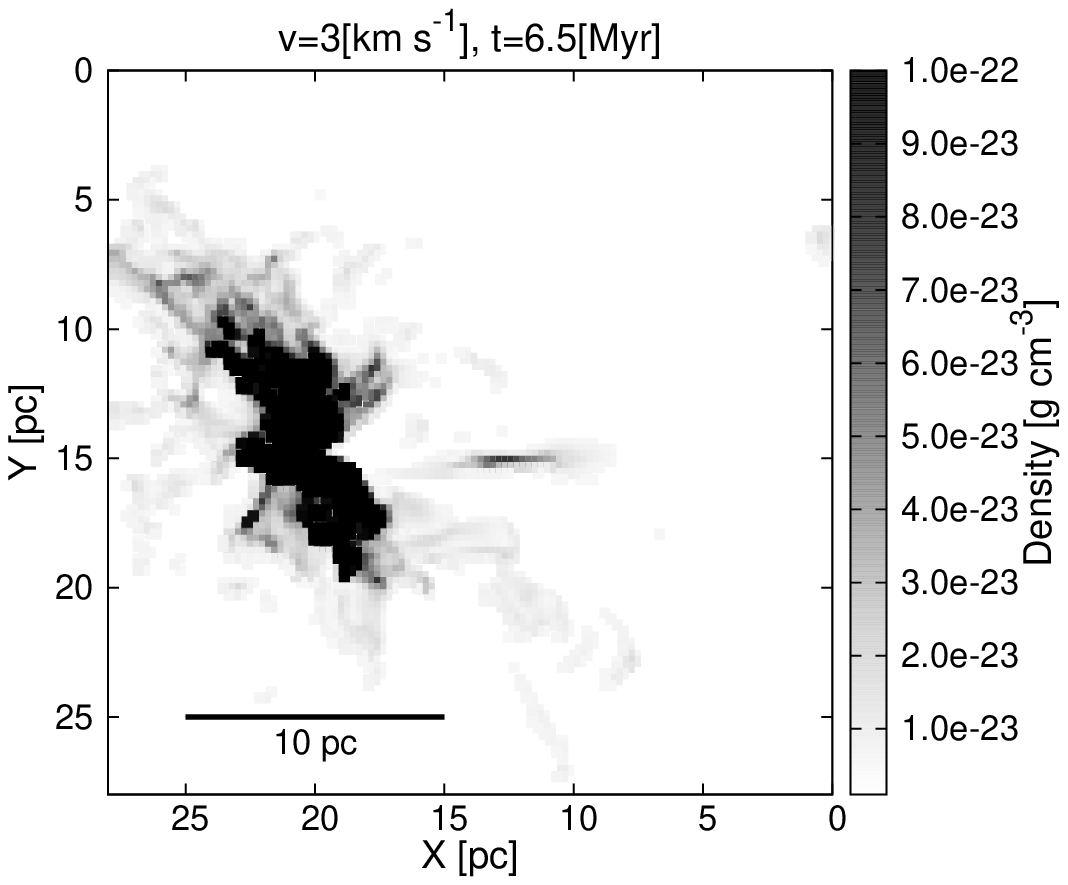}
\FigureFile(90mm,90mm){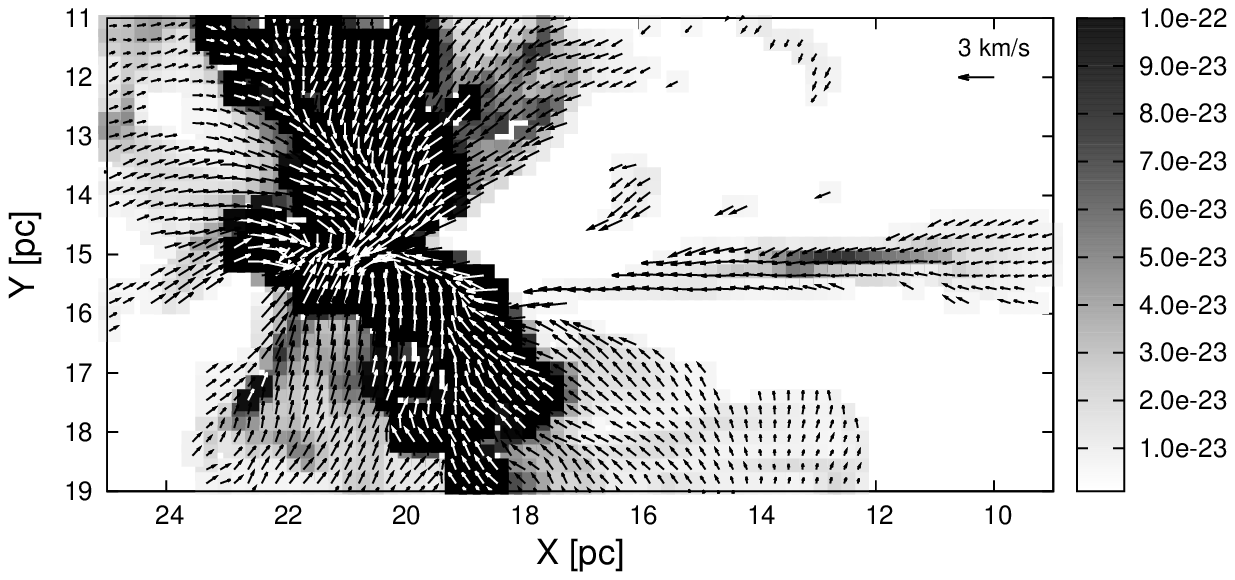}
\end{center}
\caption{(Left) CCC model: the relative velocity $3\>{\rm km\>s^{-1}}$ and timescale 6.5$\>$Myrs after collision by \citet{2014ApJ...792...63T} (the grey bar shows the density of $\atom{CO}{}{12}$ $(J\,=\,1$--$0)$ molecular cloud). (Right) CCC model by \citet{2014ApJ...792...63T}. An arc-like component (left) and a linear component (right) are detected. The vectors velocity field and the grey bar shows the density of CO molecular cloud. The upper right arrow shows $3\>{\rm km\>s^{-1}}$ scale.}\label{cloud_lb_2}
\end{figure*}

\subsubsection{Comparison with theoretical study of the cloud-cloud collision phenomenon}

As described above and shown in the left panel of figure \ref{cloud_lb}, HC$\,$5506 has structural characteristics comprising arc-like and linear structures, which is quite similar to the results of smoothed particle hydrodynamics (SPH) simulations of the cloud-cloud collision (CCC) phenomenon presented by \citet{2014ApJ...792...63T}, where two clouds collide with each other, triggering rapid gas compression. Subsequently molecular clouds can collapse faster than the free-fall time to form stars. Recent theoretical and observational studies show that CCC is one of the triggering phenomena of high-mass star formation (\cite{1992PASJ...44..203H}; \cite{2014ApJ...792...63T}; \cite{2018PASJ...70S..58T}; \cite{2014ApJ...780...36F}; \cite{2018ApJ...859..166F}; \cite{2015MNRAS.450...10H}; \cite{2015ApJ...806....7T}).

This structural similarity between the CO map of HC$\,$5506 and SPH simulation presented by \citet{2014ApJ...792...63T} implies us that IRAS$\,$01123+6430 may have formed through the CCC phenomenon. The coexistence of massive YSOs of type B1--B2 and quiescent molecular gas (HC$\,$5506) is also consistent with this scenario.

\citet{2014ApJ...792...63T} show that an arc-like structure forms in all cases that a pre-collision relative velocity between two colliding clouds is $3$, $5$, or $10\>{\rm km\>s^{-1}}$, while a linear structure forms only in cases of $3$ and $5\>{\rm km\>s^{-1}}$ relative velocities. On the basis of this, we assume that the relative velocity is likely to be in the range of $3$--$5\>{\rm km\>s^{-1}}$ at pre-collision.

Below we discuss a possible formation scenario for IRAS$\,$01123+6430 that it formed through the CCC mechanism with a relative velocity of $3$--$5\>{\rm km\>s^{-1}}$. Note that a comparison between the HC$\,$5506 (the left panel of figure \ref{cloud_lb}) and the CCC model (the left panel of figure \ref{cloud_lb_2}) indicates striking similarity within in a factor of a few parsecs in size scale.

\subsubsection{Possible formation scenario of IRAS$\,$01123+6430}

IRAS$\,$01123+6430 is located at the north-eastern end of the linear component in HC$\,$5506 as indicated with a circle in the left panel of figure \ref{cloud_lb}. The linear structure is thought to be a trail which is formed along the path of a small cloud. The left panel of figure \ref{cloud_lb_2} shows a snapshot of the gas distribution and velocity field 6.5$\>$Myrs after a CCC event occurred with a relative velocity of $3\>{\rm km\>s^{-1}}$, obtained via SPH simulations \citep{2014ApJ...792...63T}. As seen in the right panel of figure \ref{cloud_lb_2}, the gas surrounding the linear structure seems to accumulate onto the trail because of self-gravity caused by the density enhancement along the trail.

In HC$\,$5506 the separation between the arc-like and the linear structures is $7^{\prime}$, which corresponds to a physical distance of 13.5$\>$pc, which is calculated with the newly obtained trigonometric distance of 6.61$\>$kpc. Considering that the relative velocity is $3$--$5\>{\rm km\>s^{-1}}$ suggested by the model of \citet{2014ApJ...792...63T}, it is inferred that the collision of two clouds occurred 2.6--4.4$\>$Myrs ago, by dividing the physical distance by the relative velocity. This timescale of a few megayears is consistent with that of high-mass star formation.

The right panel of figure \ref{maser_dist_vect} shows that feature 4 has a proper motion vector directed south-westwards, and blue line-of-sight velocity, while feature 1 has a proper motion vector directed north-westwards, and red line-of-sight velocity. The proper motion of feature 6 is aligned with the direction of the linear structure and seems consistent with the scenario that the YSO formed in the two clouds colliding in the direction of the linear structure.

Our observational investigation of the maser source and associated molecular cloud obtained with VERA and FCRAO well match the physical signature predicted by the SPH simulation of CCC phenomenon as described above. Therefore, we conclude that IRAS$\,$01123+6430 is a good observational example of slowly colliding molecular clouds with relative velocity of $3$--$5\>{\rm km\>s^{-1}}$, which is efficient at forming massive stars, as \citet{2014ApJ...792...63T} suggested. Since the outer Galaxy is a region where the gradient of the angular motion of the Galactic rotation is small, the cloud-cloud velocity dispersion is relatively small compared to the inner Galaxy and slow CCC seems dominant. IRAS$\,$01123+6430 is located in the outer Galaxy and our case study results are consistent with the idea that the slow CCC is important for massive star formation in the outer Galaxy.

\section{Summary}

We conducted VLBI astrometric observations toward IRAS$\,$01123+6430 targeting the ${\rm H_2O}$ maser line with VERA. The annual parallax and the trigonometric distance were measured to be $0.151\,\pm\,0.042\>{\rm mas}$ and $6.61^{+2.55}_{-1.44}\>{\rm kpc}$, respectively. The systematic proper motion components were measured to be $(\mu_\alpha{\rm cos}\delta,\ \mu_\delta)\ =\ (-1.44\,\pm\,0.15,\ -0.27\,\pm\,0.16)\>{\rm mas\>yr^{-1}}$, and the 3-D velocity vectors were calculated to be $(U,\ V,\ W)\ =\ (81.2\,\pm\,14.2,\ 226\,\pm\,17,\ -7.5\,\pm\,7.3)\>{\rm km\>s^{-1}}$. Using the trigonometric distance and the galactic constants $(R_0,\ V_0)\ =\ (8.05\,\pm\,0.45\>{\rm kpc},\ 238\,\pm\,14\>{\rm km\>s^{-1}})$ adopted from \citet{2012PASJ...64..136H}, we converted those to $(V_{R_{\rm in}},\ V_{\theta},\ V_z)\ =\ (-19.1\,\pm\,5.8,\ 239\,\pm\,22,\ -7.5\,\pm\,7.3)\>{\rm km\>s^{-1}}$ in cylindrical coordinates. The bolometric luminosity of the YSO was derived to be $L_{\rm Bol}\ =\ (3.11\,\pm\,2.86)\ \times\ 10^3\>L_{\odot}$, which corresponds to spectral type B1--B2. All physical parameters, described here, are summarized in table \ref{phys_parm}.

Archival data taken from the FCRAO Outer Galaxy Survey shows that IRAS$\,$01123+6430 is associated with a molecular cloud named HC$\,$5506 which consists of arc-like and linear components. The structural characteristics of HC$\,$5506 are consistent with the predictions of SPH simulations of the slow CCC phenomenon conducted by \citet{2014ApJ...792...63T}. Observational confirmation of the coexistence of arc-like and linear structures implies that the relative velocity of the collision between the initial two clouds was $3$--$5\>{\rm km\ s^{-1}}$. Using this collisional velocity and the distance between two components (the arc-like component and the linear component), the CCC event is thought to have occurred 2.6--4.4$\>$Myrs ago. Our finding supports a hypothesis that the slow CCC phenomenon is an important trigger of massive star-formation in the outer Galaxy. 

\begin{table*}
        \caption{Summary of physical parameters for IRAS$\,$01123+6430$^{\ast}$}\label{phys_parm}
  \begin{center}
    \begin{tabular}{ll}
      \hline
      \hline
Position of Target & $(\alpha,\ \delta)\ =\ (\timeform{1h15m40.8s},\ \timeform{+64D46^{\prime}40.8^{\prime\prime}})$ (J2000) \\
 & $(l,\ b)\ =\ (\timeform{125.51D},\ \timeform{+2.03D})$ \\
LSR Velocity & $V_{\rm{LSR}}\ =\ -54.7\>{\rm km\>s^{-1}}$ \\
Annual Parallax & $\varpi\ =\ 0.151\,\pm\,0.042\>{\rm mas}$ \\
Distance from the Sun & $D\ =\ 6.61^{+2.55}_{-1.44}\>{\rm kpc}$ \\
Galactocentric Distance & $R\ =\ 13.04\,\pm\,2.24\>{\rm kpc}$ \\ 
Proper Motion & $(\mu_\alpha{\rm cos}\delta,\ \mu_\delta)\ =\ (-1.44\,\pm\,0.15,\ -0.27\,\pm\,0.16)\>{\rm mas\>yr^{-1}}$ \\
 & $(\mu_l{\rm cos}b,\ \mu_b)\ =\ (-1.40\,\pm\,0.14,\ -0.40\,\pm\,0.17)\>{\rm mas\>yr^{-1}}$ \\
 & $(v_l,v_b)\ =\ (-44.0\,\pm\,17.5,\ -12.6\,\pm\,7.3)\>{\rm km\>s^{-1}}$ \\
3-D Velocity Vectors & $(U,\ V,\ W)\ =\ (81.2\,\pm\,14.2,\ 226\,\pm\,17,\ -7.5\,\pm\,7.3)\>{\rm km\>s^{-1}}$\\
3-D Velocity Vectors (in the cylindrical coordinate) & $(V_{R_{\rm in}},\ V_{\theta},\ V_z)\ =\ (-19.1\,\pm\,5.8,\ 239\,\pm\,22,\ -7.5\,\pm\,7.3)\>{\rm km\>s^{-1}}$ \\
 Bolometric Luminosity of the YSO & $L_{\rm Bol}\ =\ (3.11\,\pm\,2.86)\ \times\ 10^3\>L_{\odot}$\\
 Spectral Type of the YSO (ZAMS) & B1--B2\\
      \hline
     \hline
    \end{tabular}
    \end{center}
{$^\ast$Physical parameters are derived using the galactic constants, $(R_0,\ \Theta_0)\ =\ (8.05\,\pm\,0.45\>{\rm kpc},\ 238\,\pm\,14\>{\rm km\>s^{-1}})$ \citep{2012PASJ...64..136H}, the standard solar motion, $(U^{\rm Std}_{\odot},\ V^{\rm Std}_{\odot},\ W^{\rm Std}_{\odot})\ =\ (10.3,\ 15.3,\ 7.7)\>{\rm km\>s^{-1}}$ \citep{1986MNRAS.221.1023K}, and peculiar solar motion, $(U_{\odot},\ V_{\odot},\ W_{\odot})\ =\ (10.0,\ 12.0,\ 7.2)\>{\rm km\>s^{-1}}$ \citep{2012PASJ...64..136H}.}
\end{table*}













\begin{ack}
We would like to thank the VERA members for supports in our observations of IRAS$\,$01123+6430. We thank Dr. H. Imai for helpful comments on our work. And we also thank Dr. T. Hirota for support in parallax fitting with the VEDA analysis system. We would also like to thank Dr. M. J. Reid for carefully reading our paper and for helpful comments to improve it. RB acknowledges support through the EACOA Fellowship from the East Asian Core Observatories Association. AH is supported by the JSPS KAKENHI Grant Number JP19K03923.
\end{ack}

\appendix 
\section*{Error propagation}
We present details on estimates of errors of the proper motion, Galactocentric distance, and 3-D velocity vectors based on the error propagation.

First, let us recall that the error of function $f$, which is a function of ($x_1,\ x_2,\ ...,\ x_i,\ ...$), can be written as
\begin{eqnarray*}
\Delta f^2=\left({\partial f\over \partial x_1}\right)^2\Delta x_1^2+\left({\partial f\over \partial x_2}\right)^2\Delta x_2^2+\cdot\cdot\cdot=\sum_i\left({\partial f\over \partial x_i}\right)^2\Delta x_i^2
\end{eqnarray*}
where $\Delta f$ is the error of $f$, and $\Delta x_i$ is the error of $x_i$.

Since ($v_l$,\ $v_b$) are functions of ($\mu$,\ $D$), the errors of $v_l$ and $v_b$ can be written as,
\begin{eqnarray*}
\Delta v_l&=&\sqrt{\left({\partial v_l\over \partial(\mu_l{\rm cos}b)}\right)^2\Delta(\mu_l{\rm cos}b)^2+\left({\partial v_l\over \partial D}\right)^2\Delta D^2}\\
&=&v_l\sqrt{\left({\Delta \mu_l{\rm cos}b\over \mu_l{\rm cos}b} \right)^2+\left({\Delta D\over D}\right)^2}
\end{eqnarray*}
and
\begin{eqnarray*}
\Delta v_b&=&\sqrt{\left({\partial v_b\over \partial(\mu_b)}\right)^2\Delta(\mu_b)^2+\left({\partial v_l\over \partial D}\right)^2\Delta D^2}\\
&=&v_b\sqrt{\left({\Delta \mu_b\over \mu_b} \right)^2+\left({\Delta D\over D}\right)^2}.
\end{eqnarray*}

Similarly since $R$ is the function of ($R_0$,\ $D$), an error of $R$ can be written as,
\begin{eqnarray*}
\Delta R=\sqrt{\left({\partial R\over \partial R_0}\right)^2\Delta R_0^2+\left({\partial R\over \partial D}\right)^2\Delta D^2}
\end{eqnarray*}
where $\partial R/\partial R_0$ and $\partial R/\partial D$ are:
\begin{eqnarray*}
{\partial R\over \partial R_0}={R_0-D{\rm cos}b{\rm cos}l\over R}
\end{eqnarray*}
and
\begin{eqnarray*}
{\partial R\over \partial D}={D{\rm cos}^2b-R_0{\rm cos}b{\rm cos}l\over R}.
\end{eqnarray*}

Since $(U,\ V,\ W)$ are functions of ($V_{\rm Helio}$,\ $v_l$, $v_b$), errors of $(U,\ V,\ W)$ can be written as,
\begin{eqnarray*}
\Delta U&=&\sqrt{\left({\partial U\over \partial V_{\rm Helio}}\right)^2\Delta V_{\rm Helio}^2+\left({\partial U\over \partial v_l}\right)^2\Delta v_l^2+\left({\partial U\over \partial v_b}\right)^2\Delta v_b^2}\\
&=&\sqrt{\Delta V_{\rm Helio}^2{\rm cos^2}l{\rm cos^2}b+\Delta v_l^2{\rm sin^2}l+\Delta v_b^2{\rm cos^2}l{\rm sin^2}b},\\
\Delta V&=&\sqrt{\left({\partial V\over \partial V_{\rm Helio}}\right)^2\Delta V_{\rm Helio}^2+\left({\partial V\over \partial v_l}\right)^2\Delta v_l^2+\left({\partial V\over \partial v_b}\right)^2\Delta v_b^2}\\
&=&\sqrt{\Delta V_{\rm Helio}^2{\rm sin^2}l{\rm cos^2}b+\Delta v_l^2{\rm cos^2}l+\Delta v_b^2{\rm sin^2}l{\rm sin^2}b},\\
\Delta W&=&\sqrt{\left({\partial W\over \partial V_{\rm Helio}}\right)^2\Delta V_{\rm Helio}^2+\left({\partial W\over \partial v_l}\right)^2\Delta v_l^2+\left({\partial W\over \partial v_b}\right)^2\Delta v_b^2}\\
&=&\sqrt{\Delta V_{\rm Helio}^2{\rm sin^2}b+\Delta v_b^2{\rm cos^2}b}.
\end{eqnarray*}

The errors of $V_{R_{\rm in}}$ and $V_\theta$ can be given by matrix formulas below:
\begin{eqnarray*}
\left[
\begin{array}{c}
\Delta V_{R_{\rm in}}\\
\Delta V_\theta \\
\end{array}
\right]=
\left[
\begin{array}{cc}
{\rm cos}\phi & -{\rm sin}\phi \\
{\rm sin}\phi & {\rm cos}\phi \\
\end{array}
\right]
\left[
\begin{array}{c}
\Delta U\\
\Delta V\\
\end{array}
\right].
\end{eqnarray*}



\newpage


\end{document}